\documentclass[preprint,12pt]{revtex4}
\usepackage{graphicx}
\usepackage{epsfig}
\usepackage{amsmath}
\usepackage{graphicx}% Include figure files
\usepackage{dcolumn}% Align table columns on decimal point
\usepackage{bm}% bold math
\usepackage{amssymb}
\usepackage{epsf}
\usepackage{subfigure}
\usepackage{subeqnarray}
\usepackage{epstopdf}
\usepackage{color}

\begin{document}

\newcommand{\pr}{\partial}
\newcommand{\rta}{\rightarrow}
\newcommand{\lta}{\leftarrow}
\newcommand{\ep}{\epsilon}
\newcommand{\ve}{\varepsilon}
\newcommand{\p}{\prime}
\newcommand{\om}{\omega}
\newcommand{\ra}{\rangle}
\newcommand{\la}{\langle}
\newcommand{\td}{\tilde}

\newcommand{\mo}{\mathcal{O}}
\newcommand{\ml}{\mathcal{L}}
\newcommand{\mathp}{\mathcal{P}}
\newcommand{\mq}{\mathcal{Q}}
\newcommand{\ms}{\mathcal{S}}

\newcommand{\nl}{$\newline$}
\newcommand{\nll}{$\newline\newline$}

\title{Quantum effects in biology: master equation studies of exciton motion in photosynthetic systems}
\author{Navinder Singh}
\affiliation{Physical Research Laboratory, Navrangpura, Ahmedabad-380009 India.}

\begin{abstract}
The present review is devoted to our recent studies on the excitonic motion in photosynthetic systems. In
photosynthesis, the light photon is absorbed to create an exciton in the antenna complex of the photosynthetic
pigments. This exciton then migrates along the chain-biomolecules, like FMO complex, to the reaction centre
where it initiates the chemical reactions leading to biomass generation.  Recently, it has been  experimentally
observed that the exciton motion is highly quantum mechanical in nature i.e., it involve long time ($\sim 600$
femto sec) quantum coherence effects. Traditional  semiclassical theories like Forrester's and second Born
master equations cannot be applied. We point out why the 2nd Born non-Markovian master  equation and its
Markovian limit (also called the Redfield master equation) cannot be used to explain the observed long
coherences. Briefly, the reason is that these approaches are perturbative in nature and in real light
harvesting systems various couplings (system-system and system-bath) are of the similar order of magnitude.
Various new approaches are being developed to go beyond the above two limiting theories. The present review is
not a review in the usual sense of the word as we summarize our own approaches and only refer to the literature
for the other ones. A brief introduction to the sophisticated 2D photon echo spectroscopy is also given at the
end with an emphasis on the underlying physics of the multidimensional echo spectroscopies. 
\end{abstract}

\maketitle


\tableofcontents

\section{Introduction to Excitation Energy Transfer (EET) and the statement of the Problem}

Photosynthesis provides chemical energy for almost all life on earth. Understanding of the natural
photosynthesis can enable us to construct artificial photosynthesis devices and a solution to the future
energy problems. We know that the safe (green) energy resources is a big challenge of the future and we know
that our present energy technology is seriously disturbing  our environment\cite{mov}. This prompts us to
investigate `` Green" resources of energy and photosynthesis is one of them. Photosynthesis is an interesting
phenomenon. On the average, on earth, biomass worth twice the mass of the Great Pyramid of Giza ($\sim10$
million tons) is being produced every hour or $\sim 3$ million kg/sec.

%------------------------------------fig No. --------------------------
\begin{figure}
\begin{tabular}{cc}
\includegraphics[height=40mm]{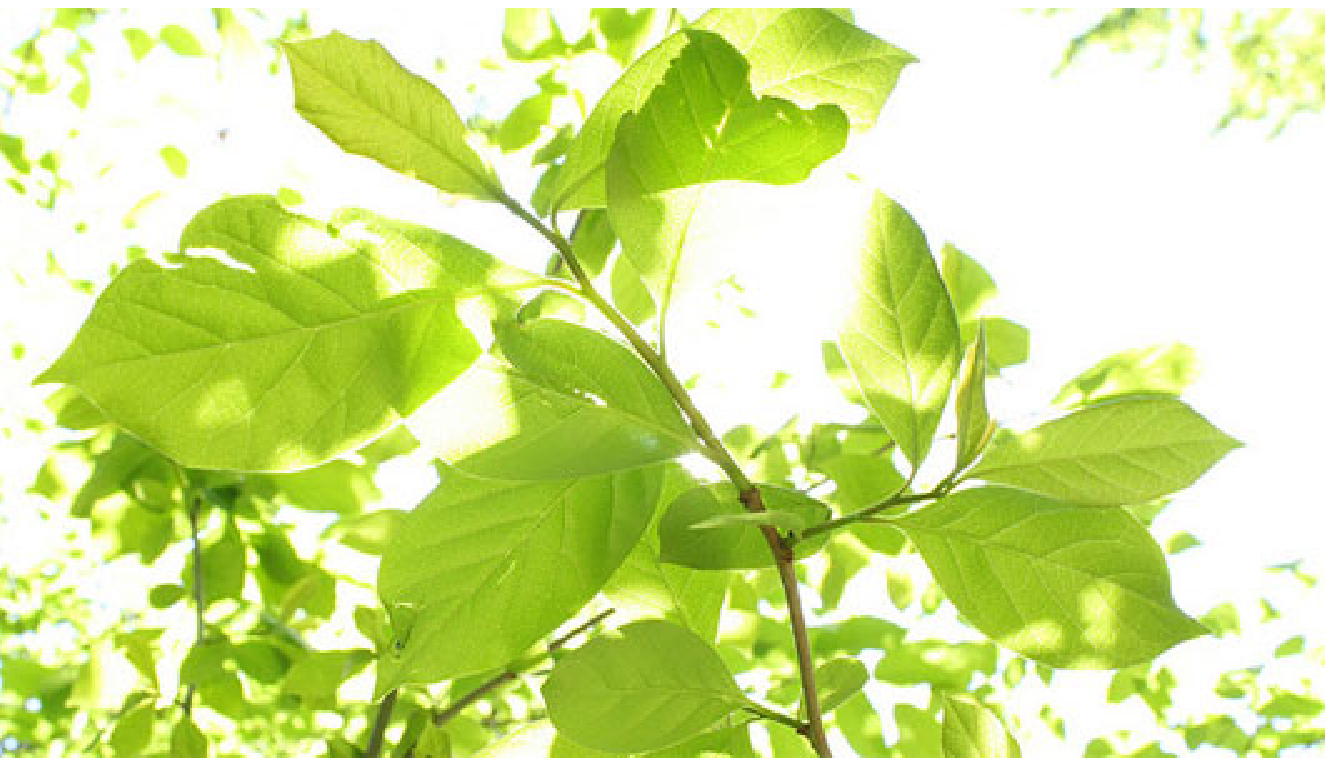}&
\includegraphics[height=40mm]{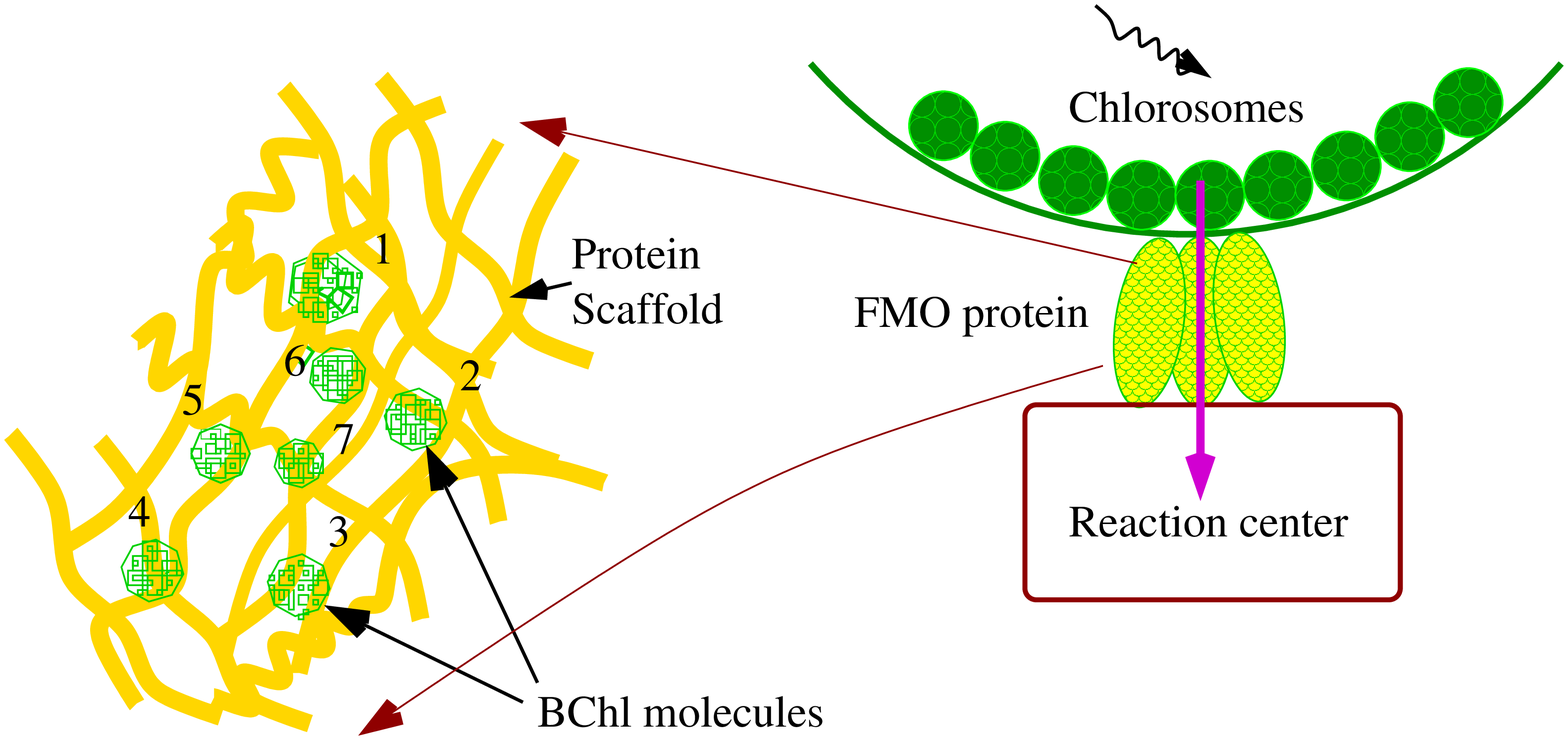}
\end{tabular}
\caption{On the average, on earth, biomass worth twice the mass of the Great Pyramid of Giza ($\sim10$ million
tons) is being produced every hour. (Right) FMO protein acts as a ``wire" which connects an antenna to a
reaction center. }\label{fig2}
\end{figure}
%--------------------------------------------------------------------

In photosynthetic systems a central role is played by the energy transport ''wire"--the FMO protein which is a
trimer made of identical subunits containing seven bacteriochlorophyll (BChl) molecules each (Fig.~\ref{fig2}).
The photosynthesis process can be divided into the following steps\cite{books1}:
\begin{enumerate}
\item A light photon is absorbed to create an excitation in the antenna complex of the 
photosynthetic pigments. 
\item This excitation then migrates along the chain-biomolecules, like FMO complex, to the
reaction center where it initiates the chemical reactions leading to biomass generation.

% %------------------------------------fig No. --------------------------
% \begin{figure}
% \includegraphics[height=50mm]{web.EPS}
% \caption{FMO protein acts as a ``wire" which connects an antenna to a reaction center.}
% \end{figure}\label{fig2}
%--------------------------------------------------------------------
\end{enumerate}

Quantum dynamics of EET (excitation energy transfer) can be very easily analyzed in the following two
limiting cases. We identify first the couplings:
\nl
{\bf Two important couplings: }
\begin{enumerate}
\item Inter BChl molecules (system-system) coupling $J$ (which is responsible for EET). 
\item The system-bath (BChl molecule-protein) coupling $\lambda$ (which is responsible for decoherence).
\end{enumerate}
\nl
{\bf Two important timescales:}
\begin{enumerate}
\item Excitation transfer time scale $\tau_{transfer}\equiv \frac{\hbar}{J} \sim 265 fs,~~for~J \sim 20 cm^{-1} 
\sim 4\times 10^{-22} joules$ (usual in photosynthesis pigments).
\item Decoherence time scale $\tau_{deco} \equiv \frac{\hbar}{\lambda}$.
\end{enumerate}

If the system-bath coupling is very weak and $\tau_{deco} >> \tau_{transfer}$, the system is
almost closed and dynamics is quantum mechanical in nature, i.e., one can use the Shroedinger's equation to
analyze it in the extreme case. But in the opposite case $\tau_{deco} << \tau_{transfer}$ (strong system-bath
coupling), the system is almost open, decoherence rate is very fast and the dynamics is almost incoherent. 
One can analyze the process with simple Pauli type master equation with rate of transfer of the excitation from
one molecule to other calculated with Fermi's golden rule (Forester's theory)\cite{for}. 

%------------------------------------fig No. --------------------------
\begin{figure}
\includegraphics[width= 12cm, height=5cm]{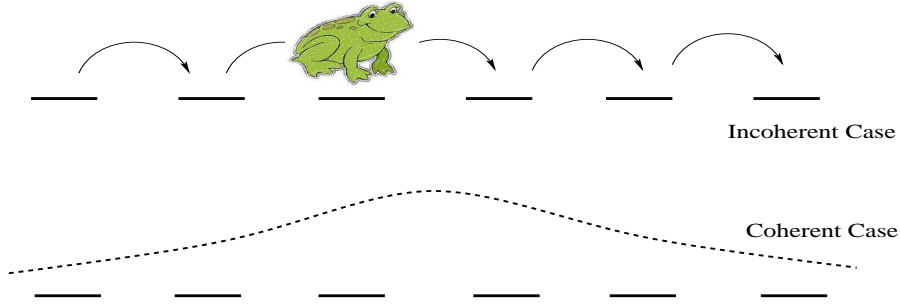}
\caption{Two extreme cases:  (1) Upper: incoherent (classical diffusive motion), 
(2) Lower: ultra-quantum (delocalized excitation). The theoretical investigation of the intermediate case is a
big challenge.}\label{ext}
\end{figure}

The important problem arises in the intermediate regime as real light harvesting systems do 
not fall in either of the extreme cases. Recently, it has been experimentally observed that
the exciton motion is highly quantum mechanical in nature i.e., it involve long time ($\sim 600$ fs) quantum
coherence effects\cite{dis}. These discoveries caused a lot of excitement in field\cite{exct} as the traditional
view was of incoherent "hopping" of the exciton(Fig.~\ref{ext}). In the weak system-bath coupling case, the
standard approach was the 2nd Born quantum master equation which is a perturbative quantum master equation (upto
second order in system-bath interaction).  Its Markovian and secular approximation is known as Redfield master
equation\cite{books2}. The 2nd Born quantum master equation can be obtained from the Nakajima-Zwanzig projection
operator technique by restricting the perturbation series upto second order\cite{books2}. These simple and
powerful projection operator techniques were introduced by Zawnzig in 1960's in then active field of
non-equilibrium statistical mechanics. As the the light harvesting pigments fall in the intermediate regime
(system-bath coupling is of the order of system-system coupling) one clearly cannot use the 2nd order
perturbative quantum master equation for its study in its original form. But in its modified form its scope
becomes wider\cite{gene}.

It is of interest to quantitatively know upto what value of system-bath coupling strength and other
important couplings in the problem, one can use the 2nd Born master equation. Section II  deals with this.

In an important case of fast bath relaxation (when bath degrees-of-freedom re-organize very fast as compared to
the transfer time scale of the exciton) a very useful approximation can be made. The details of which are given
below. This is called the Markovian approximation. In the following sections we summarize our
study\cite{nav} of the quantitative determination of the regime of validity of
the second order approximation and the Markovian approximation.

\section{Microscopic approach: 2nd Born master equation}

As is well known that the 2nd Born quantum master equation can be obtained from the
Nakajima-Zwanzig projection operator technique by restricting the perturbation series upto second
order in the system-bath interaction\cite{books2}. In the following we will apply this master equation to a
concrete model of a dimer (open two-state quantum system) which caricature the dynamics of decoherence in a
typical photosynthetic system\cite{if}. 

\subsection{Non-Markovian solution}

Projection super-operators and dynamics of the relevant system:
The total system (relevant system (electronic part) + Bath (phonons)) dynamics is pure quantum in nature. The
partial time derivative of the total density matrix is given by Liouville-van Neumann equation (classical
equivalent is the invariance of the ''extension" in phase space):

\begin{equation}
\frac{\pr \hat{\rho}_{total}^I (t)}{\pr t} = \ml(t) \hat{\rho}_{total}^I (t) \equiv -\frac{i}{\hbar}
[H^I_{el-ph} + H_b^I,\hat{\rho}_{total}^I (t)]
\end{equation}

Here the interaction representation is used $\hat{O}^I(t) = U^\dagger_S (t) \hat{O} U_S(t),~~U_S(t) =
Exp(-\frac{i}{\hbar} \hat{H}_{s} t)$ (see for details any standard refs\cite{books2}). $H_{el-ph}$ is the
system-bath interaction Hamiltonian (for the 2nd Born approximation $||H_{el-ph}||\ll||H_{s}||$. $H_b$ is the
bath Hamiltonian. To construct the equation-of-motion for the relevant part of $ \hat{\rho}_{total}^I (t)$ i.e.,
$\hat{\rho}(t)$, 
one defines the super-operator:
\begin{equation}
\mathp \hat{O} = \hat{R}_{eq} tr_{R} (\hat{O}),
\end{equation}

called the projection super-operator, one also defines $\mq = I - \mathp$. By applying $\mathp,~~\mq$ on the 
Liouville-van Neumann equation turn-by-turn, we get:

\begin{eqnarray}
&&\frac{\pr \mathp \hat{\rho}_{total}^I (t)}{\pr t} =\mathp \ml(t) (\mathp+\mq) \hat{\rho}_{total}^I (t)\nonumber\\
&&\frac{\pr \mq \hat{\rho}_{total}^I (t)}{\pr t} =\mq \ml(t) (\mathp+\mq) \hat{\rho}_{total}^I (t).
\end{eqnarray}

Solving the second equation formally for the irrelevant part  $(\mq \hat{\rho}_{total}^I (t))$, 
and inserting in the first, one obtains the required equation-of-motion for the relevant part $(\hat{\rho}(t)
\hat{R}_{eq} = \mathp \hat{\rho}_{total}^I (t) )$ (see for details\cite{books2}):

With 2nd Born approximation (i.e., by expanding the time evolution operator upto the first power in the
system-bath interaction\cite{books2}) and for a traditional dimer system\cite{if}:

\begin{eqnarray}
H_{tot} &=& H_s + H^{ph} +H^{el-ph}\nonumber\\
H_s   &=& H^{el} +H^{reorg}\nonumber\\
H^{el}  &=& \sum_{n=1}^2 \ep_n^0 |n\ra \la n| + J (|1\ra\la2| +|2\ra\la 1|)\nonumber\\
H^{reog}&=& \sum_{n=1}^2 \lambda_n |n\ra \la n|,~~~ \lambda_n = \sum_i \hbar \om_i d_{ni}^2/2\nonumber\\
H^{ph}  &=& \sum_{n=1}^2  h_n^{ph} ,~~~~ h_n^{ph} = \sum_i \hbar \om_i (p_i^2 +q_i^2)/2\nonumber\\
H^{el-ph}&=& \sum_{n=1}^2 V_n u_n,~~~~~~ V_n = | n\ra \la n|, ~~~ u_n = - \sum_i \hbar \om_i d_{ni}q_i,
\label{system}
\end{eqnarray}

master equation takes the form,

\begin{eqnarray}
&&\frac{\pr \rho^{I}(t)}{\pr t}= -\frac{i}{\hbar} \sum_{j=1}^2\la u_j \ra [V_j^{I},\rho^{I}]- \nonumber\\
&&\frac{1}{\hbar^2} \sum_{i,j=1}^2 \int_0^t d\tau (C_{ij}(t-\tau) [V_i^I(t),V_J^I(\tau) \rho^I(\tau)] - 
C_{ij}^*(t-\tau) [V_i^I(t),\rho^I(\tau) V_J^I(\tau)]) \label{master}
\end{eqnarray}

Here in the Hamiltonian, $| n\ra$ represents the state in which ONLY $n$th site is excited and all others are in
the ground state i.e., $|n\ra = | \phi_{n,e} \ra|\phi_{m \neq n,g}\ra$. $H_s$ is the system Hamiltonian
which consists of $H^{el}$ the electronic Hamiltonian for the two level system, and $H^{reorg}$ the
Hamiltonian for the re-organization energy (the elastic energy related to the physical organization of the bath
degrees-of-freedom). $H^{ph}$ is the phonon Hamiltonian and $H^{el-ph}$ is the system-bath coupling Hamiltonian.
In the absence of phonons, $\ep_n^0$ is the excited electronic energy of
$n^{th}$ site and $J$ is the electronic coupling between
the sites which is responsible for excitation transfer. The ground state energies
of both donor and acceptor are set equal to zero and $\lambda_j$ is
the re-organization energy of the $j^{th}$ site (Dissipated energy
in the bath after the electronic transition occurs).  $d_{ji},q_{i}, p_i$  are the dimensionless displacement of
the equilibrium configuration of the $i^{th}$ phonon mode, dimensionless coordinates, momenta of the
$i^{th}$ phonon mode respectively.

In the master equation, the bath correlation functions (bath is assumed to be a continuum of harmonic
oscillators (valid when an-harmonic terms are not important)) are:

\begin{equation}
C_{ij}(t)\equiv \la u_i(t)u_j(0)\ra - \la u_i\ra \la u_j\ra
\end{equation}

We consider a case where the characteristics of the bath as seen by both the sites are the same, 
and there is no systematic bath correlations between the sites. Thus the  bath correlation function takes the
form: $C_{ij}(t) = C(t) \delta_{ij}$:

\begin{equation}
C(t) = \int_{-\infty}^{+\infty} \frac{d\omega}{2\pi} C(\omega) e^{-i \omega t}.
\end{equation}

\begin{equation}
C(\omega) = 2 \hbar (1+ n(\omega)) J(\omega),~~~~~J(\omega) = 2 \lambda \frac{\omega \gamma}{\omega^2 + \gamma^2}
\end{equation}

Assuming the Drude-Lorentz model\cite{if} for the bath spectral density, and assuming the high temperature
approximation ($\frac{\hbar \omega}{k_B T} << 1$), as appropriate for the FMO problem, we obtain

\begin{equation}
C(t)= \frac{2\lambda}{\beta} e^{-\gamma t},~~~ \beta = \frac{1}{k_B T}
\end{equation}

\subsection{Representations}

The equation (\ref{master}) is an operator equation and this can be expressed in site or in energy
representation. In site representation, with definitions $x(t)\equiv
\rho_{11}(t)\equiv \la 1 |\hat{\rho}(t)|1\ra$ (site),
$y_1(t) \equiv \textrm{Re}[\rho_{12}(t)]$, and
$y_2(t)\equiv \textrm{Im}[\rho_{12}(t)]$, and with lengthy but straightforward calculations (see for
details\cite{nav}), the equation (\ref{master}) can be written explicitly as a set of coupled
integro-differential delay equations:
\begin{eqnarray}
 \frac{dx(t)}{dt} = && -2\frac{J}{\hbar} y_2(t)\nonumber\\
\frac{dy_1(t)}{dt}= &&\frac{\Delta}{\hbar}y_2(t) - \frac{4\lambda}{\beta\hbar^2}e^{-\gamma t} \int_0^t d\tau
e^{\gamma \tau} \nonumber\\
&&\left[\eta_1 \cos(E_{12}(t-\tau))y_1(\tau) +\eta_2 \sin(E_{12}(t-\tau))y_2(\tau) \right]\nonumber\\
\frac{dy_2(t)}{dt}=&&-\frac{\Delta}{\hbar}y_1(t) -\frac{J}{\hbar}(1-2x(t)) - \frac{4\lambda}{\beta\hbar^2}
e^{-\gamma t} \int_0^t d\tau e^{\gamma \tau}\nonumber\\
&& \left[-\eta_2 \sin(E_{12}(t-\tau))y_1(\tau) +\eta_3
\cos(E_{12}(t-\tau))y_2(\tau)+2 \Omega y_2(\tau) \right]\nonumber
\end{eqnarray}
with $ \eta_1= 1,~~~\eta_2 = -\frac{\Delta}{\sqrt{\Delta^2 +4
J^2}},~~~\eta_3=\frac{\Delta^2}{\Delta^2 + 4 J^2},~~~~
E_{12}=(E_1-E_2)/\hbar = -\frac{\sqrt{\Delta^2 +4 J^2}}{\hbar},~~~\Omega=
\frac{2 J^2}{\Delta^2 + 4 J^2}$.

For energy representation we need the eigensystem of the Hamiltonian. Let the kets $|e_{1,2}\ra$ be the
eigenstates of the Hamiltonian $H_{s}$. The reduced density matrix in energy representation can be expressed as:
\begin{equation}
\rho_{ab}^e(t) \equiv \la e_a|\hat{\rho}(t)|e_b\ra,
\end{equation}
with time evolution given as,
\begin{eqnarray}
 \frac{d\rho_{ab}^e(t)}{dt} = && - i \omega_{ab}\rho_{ab}^e
- \frac{1}{\hbar^2} \sum_{i,c,d =1}^2\int_0^t d\tau C(t-\tau) \nonumber \\
&& \left[V_i^{ac}V_i^{cd}e^{-i \omega_{cb}(t-\tau)}\rho_{db}^e(\tau)- V_i^{ac}V_i^{db}e^{-i 
\omega_{ad}(t-\tau)}\rho_{cd}^e(\tau) \right]\nonumber\\
&&- C^*(t-\tau)\left[V_i^{ac}V_i^{db}e^{-i
\omega_{cb}(t-\tau)}\rho_{cd}^e(\tau)- V_i^{cd}V_i^{db}e^{-i
\omega_{ad}(t-\tau)}\rho_{ac}^e(\tau) \right], \label{eq17}
\end{eqnarray} 
with $\omega_{ab} = (E_a - E_b)/\hbar$ and
\begin{equation}
V_1^{ac} = \frac{\alpha_a\alpha_c}{\sqrt{\alpha_a^2
+1}\sqrt{\alpha_c^2+1}},~~~ V_2^{ac} = \frac{1}{\sqrt{\alpha_a^2
+1}\sqrt{\alpha_c^2+1}}. \label{eq18}
\end{equation}

Eigensystem of the  Hamiltonian:  Assuming  $\lambda_1  =  \lambda_2 \equiv \lambda$ the
eigenvalues $E_i$ and eigenvectors $|e_i\ra$ of the system Hamiltonian
\begin{equation}
H_s = \sum_{n=1}^2 (\ep_n^0 +\lambda_n) |n\ra \la n| + J
(|1\ra\la2| +|2\ra\la 1|) \nonumber
\end{equation}
can be easily obtained as
\begin{eqnarray}
&&E_{1,2} = \frac{1}{2}(\ep_1^0 +\ep_2^0 + 2\lambda \mp \sqrt{(\ep_1^0 +\ep_2^0 + 2\lambda)^2 - 4 
(\ep_1^0\ep_2^0 - J^2 + \lambda(\ep_1^0+\ep_2^0) + \lambda^2)})\nonumber\\
&& |e_1\ra = \frac{1}{\sqrt{\alpha_1^2 +1}}\binom{\alpha_1}{1},~~~ |e_2\ra = \frac{1}{\sqrt{\alpha_2^2 +1}}
\binom{\alpha_2}{1} \nonumber\\
&&\alpha_{1,2} = \frac{1}{2 J} (\Delta \mp \sqrt{\Delta^2 + 4
J^2}),~~~\Delta = \ep_1^0-\ep_2^0. \nonumber 
\end{eqnarray}
Here the column vectors denote components in the site basis. The eigenkets are normalized and are orthogonal
$\alpha_1\alpha_2 = -1$.

\subsection{Numerical Approach}

It is well known that the numerical propagation of integro-differential equations is an involved task and time
consuming. In the following we construct a simple method to the solution of numerical integration\cite{nav}.
Specifically we utilize the exponential nature of the bath correlation function which helps to convert the set
of coupled integro-differential equations to a bigger set of ordinary differential equations: 
\begin{eqnarray}
&& f_1(t')\equiv \int_0^{t'} d\tau' e^{\tau'} \left[ \cos[\frac{E_{12}}{\gamma}(t'-\tau')] \td{y}_1(\tau') + 
\eta_2 \sin[\frac{E_{12}}{\gamma}(t'-\tau')]\td{y}_2(\tau') \right],\nonumber\\
&&f_2(t')\equiv \int_0^{t'} e^{\tau'}\td{y}_2(\tau')d\tau',\nonumber\\
&&f_3(t')\equiv \int_0^{t'} d\tau' e^{\tau'} \left[ -\eta_2 \sin[\frac{E_{12}}{\gamma}(t'-\tau')]
\td{y}_1(\tau') 
+ \eta_3 \cos[\frac{E_{12}}{\gamma}(t'-\tau')]\td{y}_2(\tau') \right].
\end{eqnarray}
With $t' = \gamma t,~~\tau' = \gamma \tau'$. Here,~$\td{y}_1(t') \equiv y_1(t'/\gamma),~~ \td{y}_2(t') \equiv
y_2(t'/\gamma)$ and we also define $\td{x}(t') \equiv
x(t'/\gamma)$.

We obtain a set of coupled ordinary differential equations (note that these are much simpler to solve as 
compared to coupled integro-differential equations):
\begin{eqnarray}
&&\dot{\td{x}}(t') = -\frac{2 J}{\gamma \hbar} \td{y}_2(t'),\nonumber\\
&& \dot{\td{y}}_1(t') = \frac{\Delta}{\gamma \hbar}\td{y}_2(t') - \frac{4 \lambda}{\beta \gamma^2 \hbar^2} 
e^{-t'} f_1(t'),\nonumber\\
&& \dot{\td{y}}_2(t') = -\frac{\Delta}{\gamma \hbar}\td{y}_1(t')-\frac{J}{\gamma \hbar} + 2\frac{J}{\gamma
\hbar} 
\td{x}(t') - \frac{8 \lambda}{\beta \gamma^2\hbar^2} \Omega e^{-t'} f_2(t') - \frac{4 \lambda}{\beta \gamma^2
\hbar^2} e^{-t'} f_3(t'),\nonumber\\
&& \ddot{f}_1(t') - e^{t'} \dot{\td{y}}_1(t') = e^{t'}\td{y}_1(t') + \frac{E_{12}}{\gamma} e^{t'}\eta_2
\td{y}_2(t') 
- \left(\frac{E_{12}}{\gamma}\right)^2 f_1(t'),\nonumber\\
&& \dot{f}_2(t') = e^{t} y_2(t),\nonumber\\
&& \ddot{f}_3(t') - e^{t'} \eta_3 \dot{\td{y}}_2(t') = e^{t'}
\eta_3\td{y}_2(t') -\frac{E_{12}}{\gamma} \eta_2{\gamma}
e^{t'}\td{y}_1(t') - \left(\frac{E_{12}}{\gamma}\right)^2 f_3(t')~
,\label{eq20}
\end{eqnarray}

Our aim is to use this to establish the parameter range over which the Markovian approximation is valid.
Before doing so we compare this method  with the traditional method of solution of
integro-differential equations\cite{vol}. In the straightforward numerical method (traditional method) the 
integro-differential equation is first written as integral equation with double
integration as $d x(t)/d t = \int_0^t f(x(t-\tau),
t) d\tau$ converted to $ x(t) = \int_0^t dt' \int_0^{t'} d\tau
f(x(t'-\tau), t') $. The double integration is then done self-consistently with numerical integration\cite{vol}.

\begin{figure}
\centering
\begin{tabular}{cc}
\includegraphics[height = 4cm, width = 6cm]{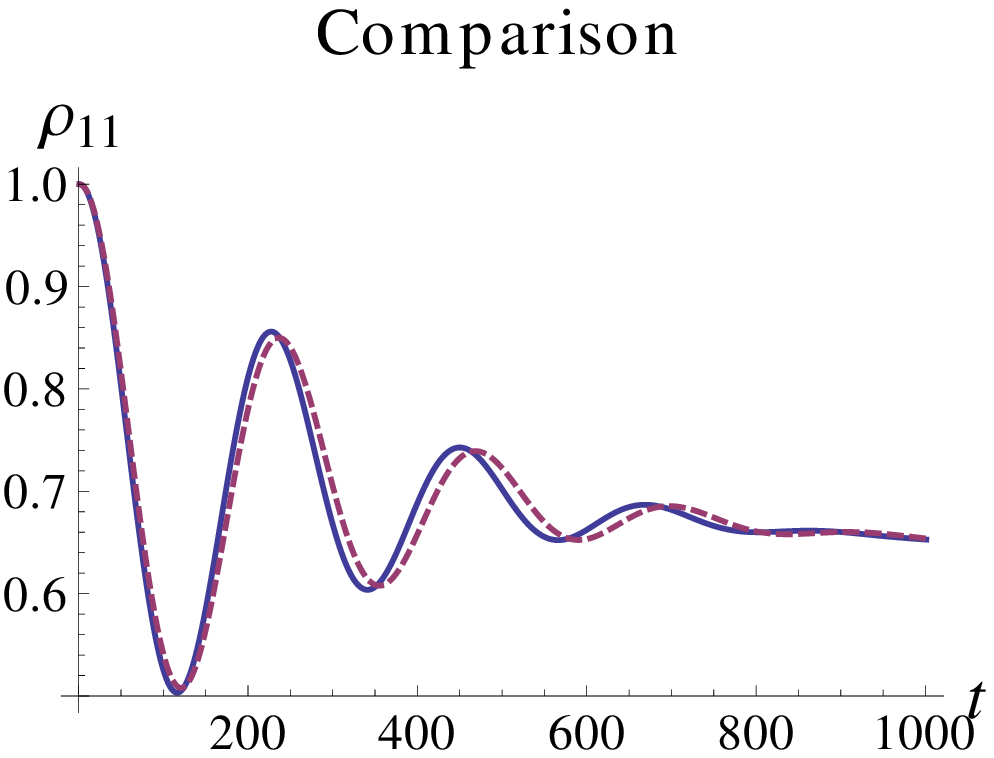}~~~~~~~&
\includegraphics[height = 4cm, width = 6cm]{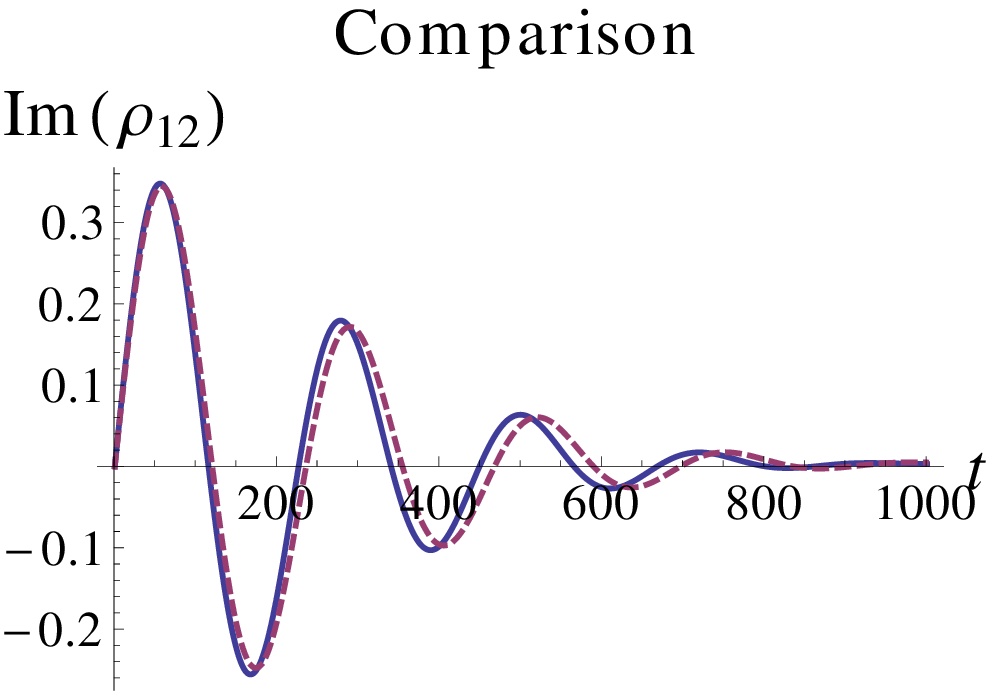}
\end{tabular}
\caption{Comparison of numerical results for a traditional method (blue-solid curve), to that introduced here 
(red-dotted curve). Parameters used are: $\gamma  =
10^{13}~\textrm{s}^{-1},~~\lambda= 2~\textrm{cm}^{-1},~~J= 50~\textrm{cm}^{-1},~~\Delta = 100~\textrm{cm}^{-1}$.
Abscissa is in fs (fs $\equiv$ femto seconds).}\label{comp}
\end{figure}

Speed: On the Lenovo ThinkCentre-i7, the traditional method took $\sim 15$ minutes to obtain the result, 
whereas the present method  took only about a few milliseconds. A sample comparison is given in
Fig.~(\ref{comp}).

%------------------------------------fig. --------------------------

\subsection{Markovian limit}\label{section4}

As mentioned before in an important case of fast bath relaxation (when bath degrees-of-freedom re-organize very
fast as compared to the transfer time scale of the exciton) a very useful approximation can be made. This is
called the Markovian approximation\cite{note100}. To do this approximation,  note that it
is particularly simple to invoke it in the energy representation (as one can ``average out" the fast
oscillations in the system's density matrix and can compare the temporal envelope of the system's density matrix
with time decay of the bath correlation function). Hence, below we first utilize the energy basis and then
convert the result back to the site representation.

The Markov approximation can be performed when the time scale on
which the envelope of the density matrix decays is much longer
than the decay time of the phonon correlation function \cite{books2}.
One can then introduce the following approximation:
\begin{equation}
\rho_{ab}^e(t-\tau) \equiv e^{- i \omega_{ab}
(t-\tau)}\td{\rho}_{ab}^e(t-\tau) \simeq e^{- i \omega_{ab}
(t-\tau)}\td{\rho}_{ab}^e(t) = e^{i \omega_{ab} \tau}
\rho_{ab}^e(t). \nonumber\end{equation}

To do this energy representation equations are first converted to dimensionless form with $\tau' = \gamma
\tau$ (see for details\cite{nav}). Putting $t-\tau = \tau'$ in the resulting
equations and then implementing the above approximation on the density matrix elements allows the time
integration to be performed easily for the case of exponential
phonon correlation function. The result is the set of Markovian equations:
\begin{eqnarray}
&&\dot{\td{\rho}}_{ab}^e(t') = - i \bar{\omega}_{ab} \td{\rho}_{ab}^e(t') \nonumber\\
&&-\frac{2\lambda}{\beta \hbar^2\gamma^2} \sum_{i,c,d} \left( \frac{V_i^{ac} V_i^{cd}}{1- i \bar{\omega}_{dc}}
\td{\rho}_{db}^e(t') - \frac{V_i^{ac} V_i^{db}}{1+ i \bar{\omega}_{db}}\td{\rho}_{cd}^e(t') \right) \nonumber\\
&& + \frac{2\lambda}{\beta \hbar^2\gamma^2} \sum_{i,c,d} \left( \frac{V_i^{ac} V_i^{db}}{1- i \bar{\omega}_{ca}}
\td{\rho}_{cd}^e(t') -
 \frac{V_i^{cd} V_i^{db}}{1+ i \bar{\omega}_{cd}}\td{\rho}_{ac}^e(t') \right) \nonumber.
\end{eqnarray}
Here $\rho^e_{ab}(t'/\gamma) \equiv \td{\rho}^e_{ab}(t'), ~~
\bar{\omega}_{ab} \equiv \omega_{ab}/\gamma$.  The results can then be transformed back to the site
representation using $\rho_{ij}(t) = \la i|{\rho}(t)|j\ra = \sum_{a,b}\la i|e_a\ra
\rho_{ab}^e \la e_b|j\ra$.

\subsection{Limiting Cases: Analytical Results}

Markovian and non-Markovian results were obtained computationally and compared for various regimes. Before
presenting them we show some interesting analytic results in two extreme cases on the parameter dependence
of the region of validity of the Markov approximation:

\subsubsection{Strong Coupling Case: $J \gg \Delta$}
For $J \gg \Delta$, we have $\alpha_1 \simeq 1,~~\alpha_2 \simeq -1,$ and $V_1^{ij} \simeq 1/2$ for $i = j$ 
and $\simeq -1/2$ for $i \ne j~~(\{i,j\} = 1,2)$ and $V_2^{i,j} \simeq 1/2$ for all $i,j$. One can then
analytically solve the coupled equations to obtain the simple expression
\begin{equation}
\tilde{\rho}_{11}^e(t') = \frac{1}{2} (e^{- \frac{4\lambda}{\beta(4 J^2 + \hbar^2 \gamma^2)} t'} + 1),
\end{equation}
for the traditional initial conditions $\tilde{\rho}_{11}^e(t'=0) =1,~~ \tilde{\rho}_{12}^e(t'=0) = 
\tilde{\rho}_{21}^e(t'=0) = 0$. The Markov approximation can be performed when the time scale on which the
envelope of the density matrix decays is much longer than the decay time of the phonon auto-correlation
function. Hence, $ \boxed{\frac{4\lambda}{\beta(4 J^2 + \hbar^2 \gamma^2)} \ll 1}$ must hold for the Markov
approximation to be valid in the $J \gg \Delta$ domain.

\subsubsection{Weak Coupling Case: $J \ll \Delta$}
\begin{figure}[htbp]
\centering
\begin{tabular}{cc}
\includegraphics[height = 4cm, width = 6cm]{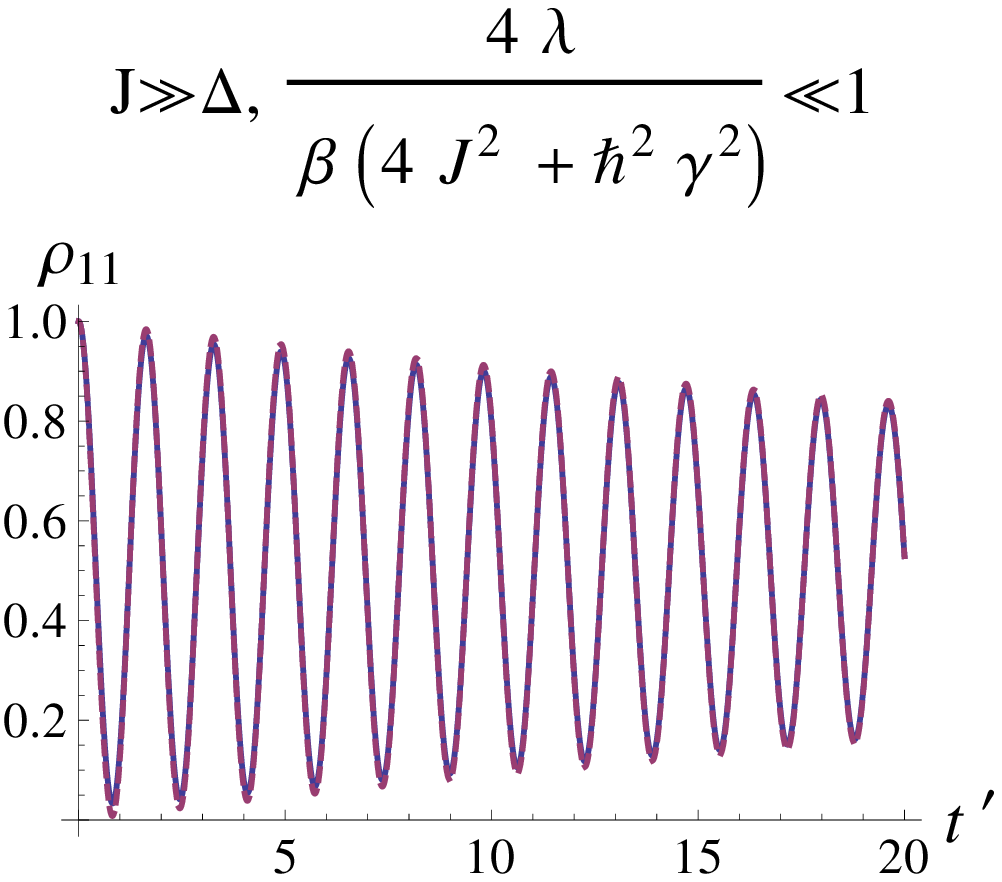}&
\includegraphics[height = 4cm, width = 6cm]{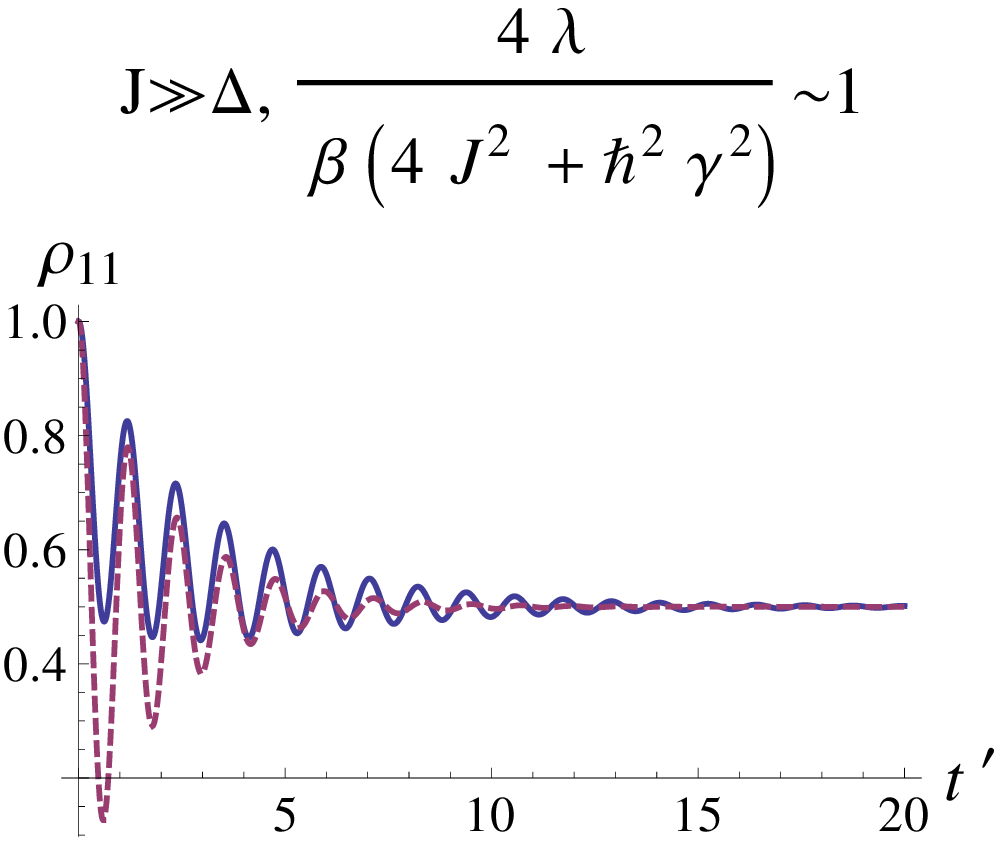}\\
\includegraphics[height = 4cm, width = 6cm]{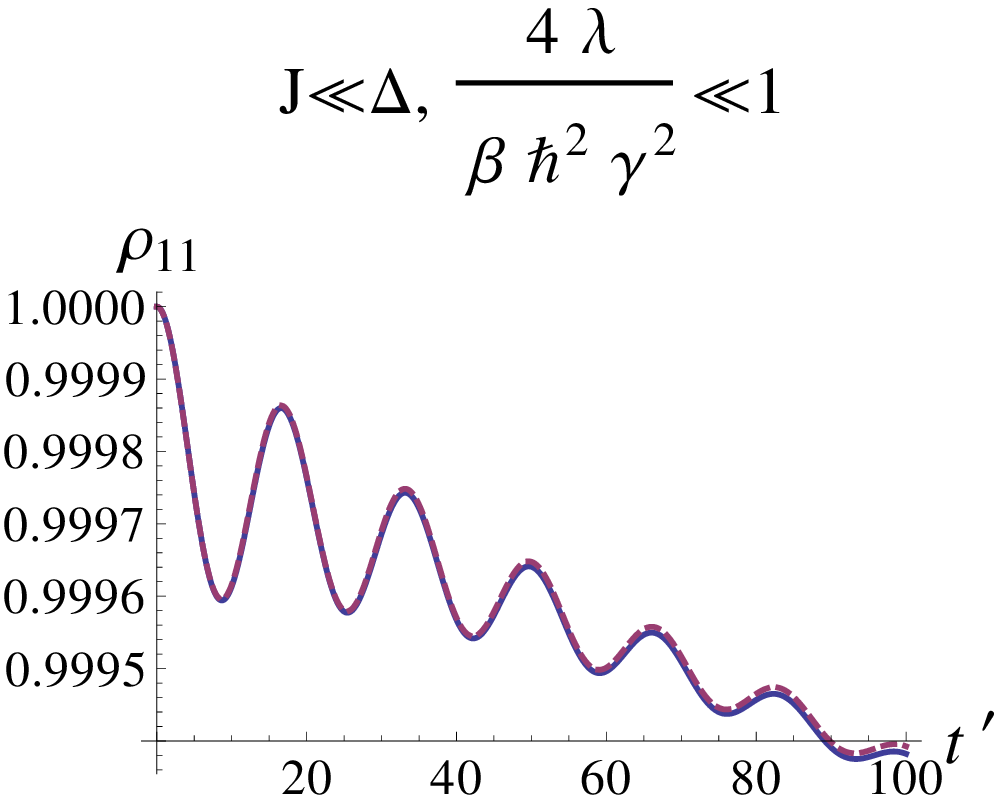}&
\includegraphics[height = 4cm, width = 6cm]{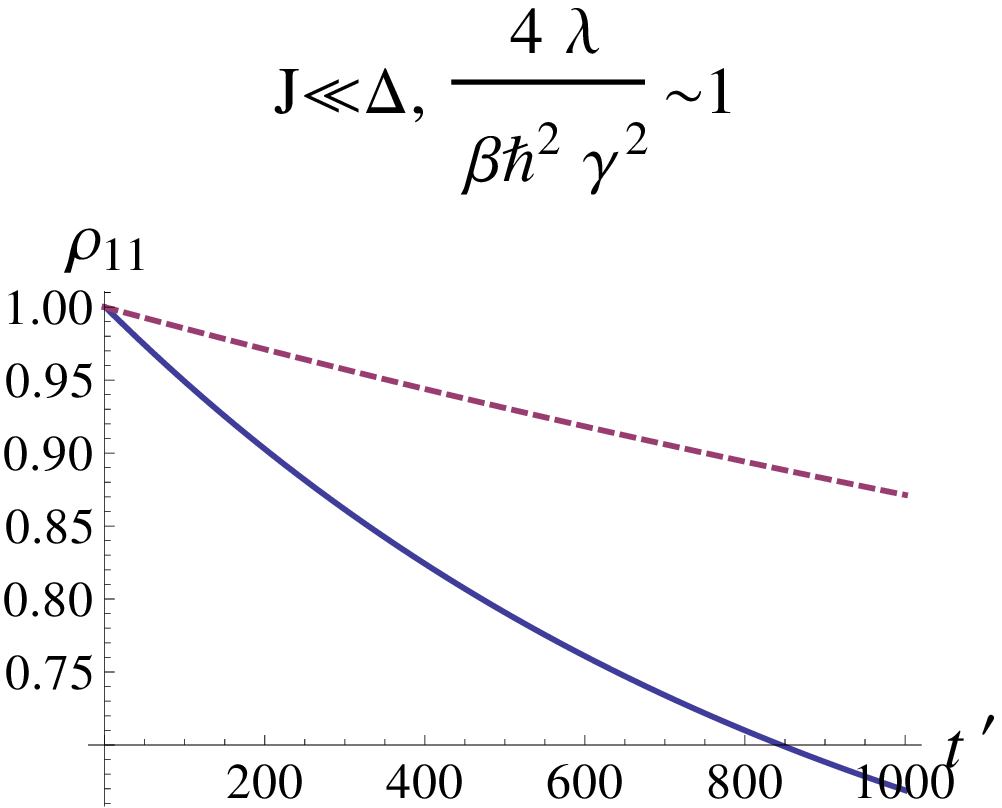}\\
\end{tabular}
\caption{Sample verification of the analytic inequalities (Table I). Time evolution of population on site 1: 
blue (solid) curve is the non-Markovian solution and red (dotted) curve is the
Markovian approximation.} \label{veri}
\end{figure}

For this case, we have $\alpha_{1,2}= \frac{1}{2J}(\Delta \mp \sqrt{\Delta^2 + 4J^2}) \simeq \frac{\Delta}{2
J}(1\mp1)$. Hence, in this domain $\alpha_1 \simeq 0$, and $\alpha_2 \simeq \Delta/J$. This leads to $V_1^{11}
=V_1^{12}=V_1^{21} \simeq 0,~~V_1^{22}\simeq 1$. $V_2^{11} \simeq 1,~~V_2^{12}=V_2^{21}\simeq J/\Delta$, and
$V_2^{22} = (J/\Delta)^2$.
\begin{eqnarray}
&&\dot{\tilde{\rho}}_{11}^e(t')= \nonumber\\
&&\frac{2/\lambda}{\hbar^2\beta \gamma^2}\left(2(J/\Delta)^2 (\Gamma +\Gamma^\ast) - 4(J/\Delta)^2 
(\Gamma +\Gamma^\ast) \tilde{\rho}_{11}^e(t') + 2 (J/\Delta) (\tilde{\rho}_{12}^e(t') + \tilde{\rho}_{21}^e(t'))
\right), \nonumber\\
&& \dot{\tilde{\rho}}_{12}^e(t') = \frac{ i \Delta}{\hbar \gamma}\tilde{\rho}_{12}^e(t')+\nonumber\\
&& + \frac{4\lambda}{\hbar^2\beta \gamma^2}\left((J/\Delta)\Gamma^\ast (2 \tilde{\rho}_{11}^e(t') -1) - 
(1 + 2 \Gamma (J/\Delta)^2) \tilde{\rho}_{12}^e(t') + 2 (J/\Delta)^2 \Gamma^\ast \tilde{\rho}_{21}^e(t')\right).
\nonumber
\end{eqnarray}
where $\Gamma = \frac{1}{1+ i\frac{\Delta}{\hbar\gamma}}$. See for details the second paper in\cite{nav}. By
separating real and imaginary parts as $\tilde{\rho}_{12}^e(t') = x(t') + i y(t')$ and writing
$\tilde{\rho}_{11}^e(t') = r(t') $, we have:
\begin{eqnarray}
&&r(t') = \frac{1}{\eta^2 + \xi^2} (\eta^2 +\xi^2 +[a\xi-b\eta]\ep \xi \nonumber\\
&& [b\eta-a\xi] \ep\xi \cos(\eta t')e^{-\xi t'} + [a\eta + b\xi] \ep\xi\sin(\eta t')e^{-\xi t'}),\nonumber\\
&& x(t') = e^{-\xi t'} (a \cos(\eta t')-b\sin(\eta t')),\nonumber\\
&&y(t') = e^{-\xi t'} (a \sin(\eta t') + b \cos(\eta t')). \label{eq45}
\end{eqnarray}
with initial conditions $r(t'=0)=1,~~x(t'=0)=a,~~y(t'=0)=b$. Here $\xi = {4\lambda}/(\hbar^2 \beta \gamma^2),~~
\eta = {\Delta}/{\hbar\gamma},$ and
$\ep = {J}/{\Delta}$. Thus, for the Markov approximation to hold requires $\boxed{\xi= 4\lambda/\beta \hbar^2
\gamma^2 \ll 1}$. These are summarized in Table~\ref{table1}. A numerical verification of these analytical
inequalities is given in Fig.~(\ref{veri}).

\begin{table}[h!]
\caption{Regimes of validity of the Markov approximation}
\begin{tabular}{lr}
Case & Markovian approximation \\\hline
$J>>\Delta$ & $\frac{4\lambda}{\beta (4 J^2 + \hbar^2 \gamma^2)}<<1 $\\\hline
$J<< \Delta$  & $\frac{4 \lambda}{\hbar^2 \beta \gamma^2}<<1$\\\hline
\end{tabular}\label{table1}
\end{table}
% %%%-----------------------------
% \begin{table}
% \begin{tabular}{lr}
% \hline
% Case & Approx. matrix elements\\\hline
% $J>>\Delta$ & $ V_k^{ij} \simeq (-1)^{k(i+j)}\frac{1}{2},~~i,j,k=1,2$\\\hline
% $J<< \Delta$  & $~V_k^{ij}\simeq \delta_{k1}\delta_{i2}\delta_{j2} +\delta_{k2} (\delta_{i1}\delta_{j1} +
%(\frac{J}{\Delta})^2\delta_{i2}\delta_{j2} + \frac{J}{\Delta}(\delta_{i1}\delta_{j2}+\delta_{i2}\delta_{j1}))$
% \\\hline
% \end{tabular}
% \end{table}

%------------------------------------fig.SET No.1 --------------------------

\subsection{Computational Results}\label{section5}

\begin{figure}[htbp]
\centering
\begin{tabular}{ccc}
\includegraphics[height = 3cm, width = 4cm]{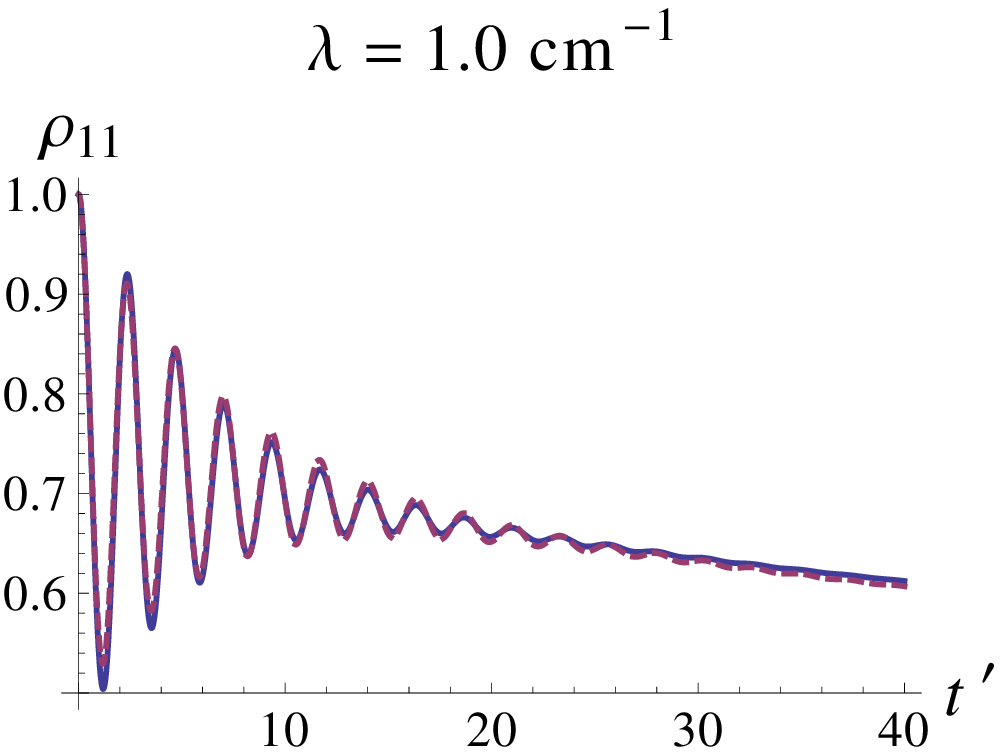}&
\includegraphics[height = 3cm, width = 4cm]{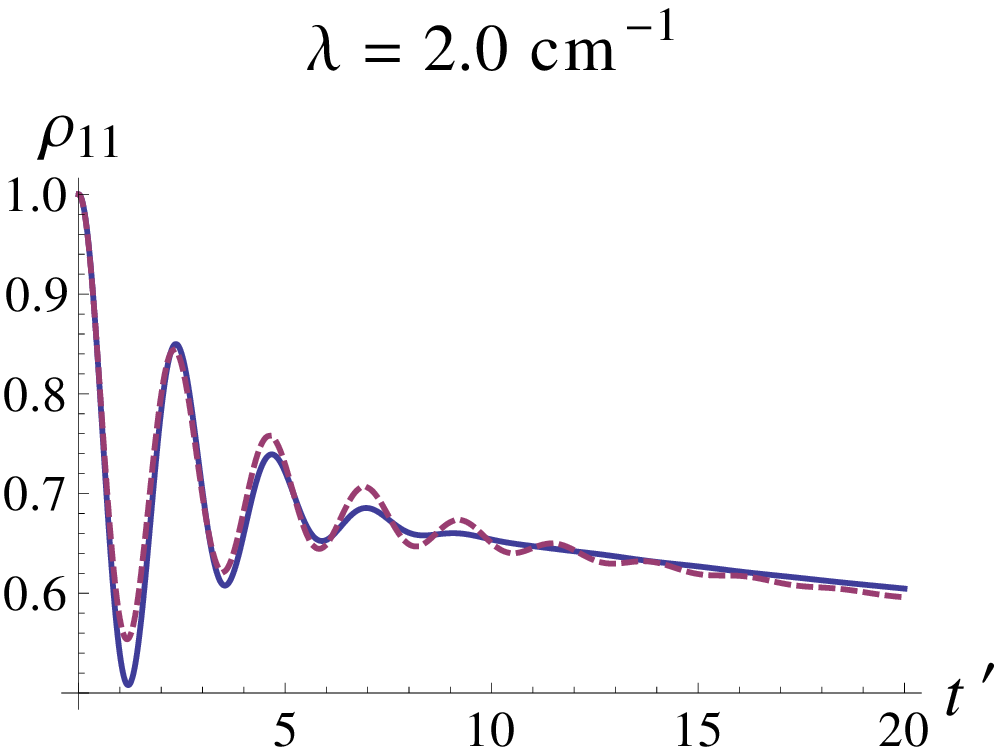}&
\includegraphics[height = 3cm, width = 4cm]{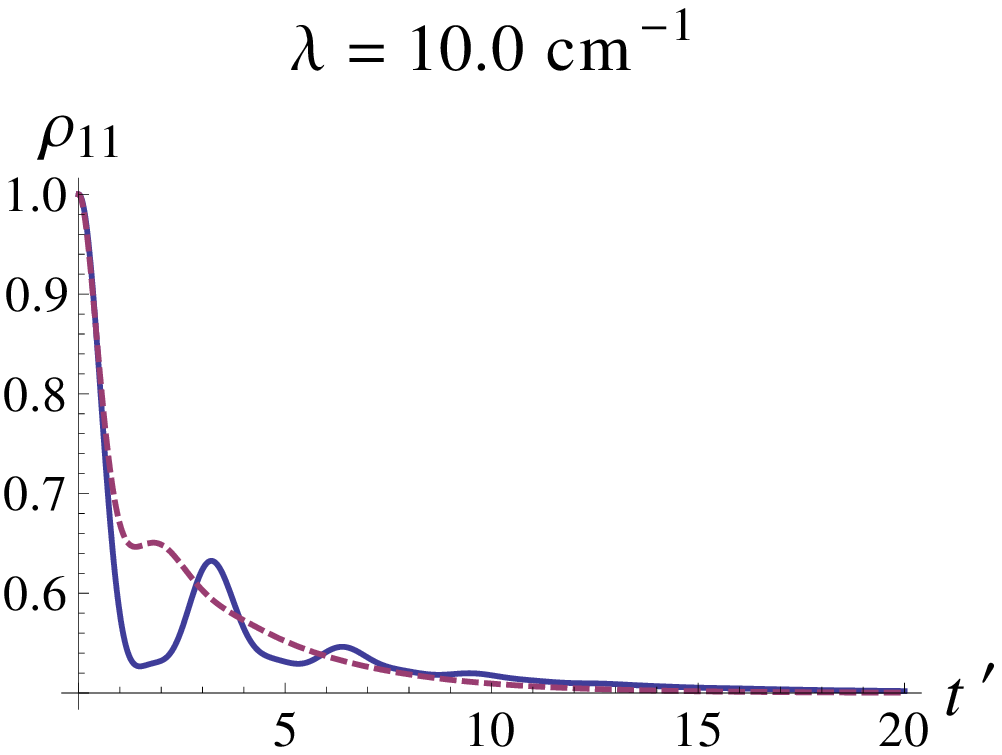}\\
\includegraphics[height = 3cm, width = 4cm]{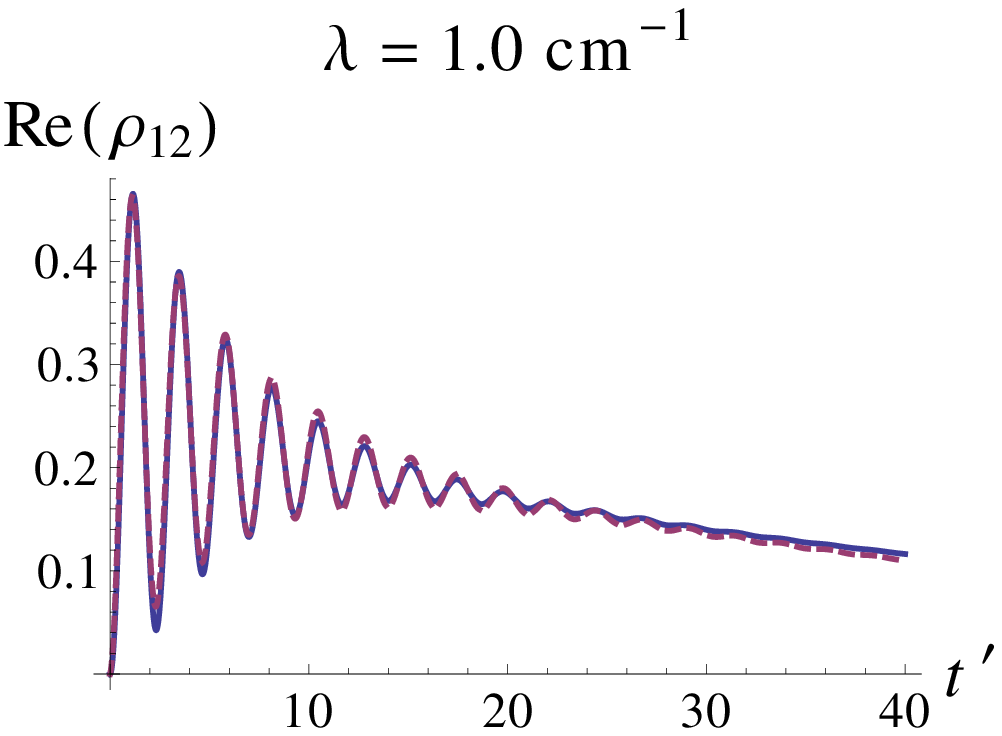}&
\includegraphics[height = 3cm, width = 4cm]{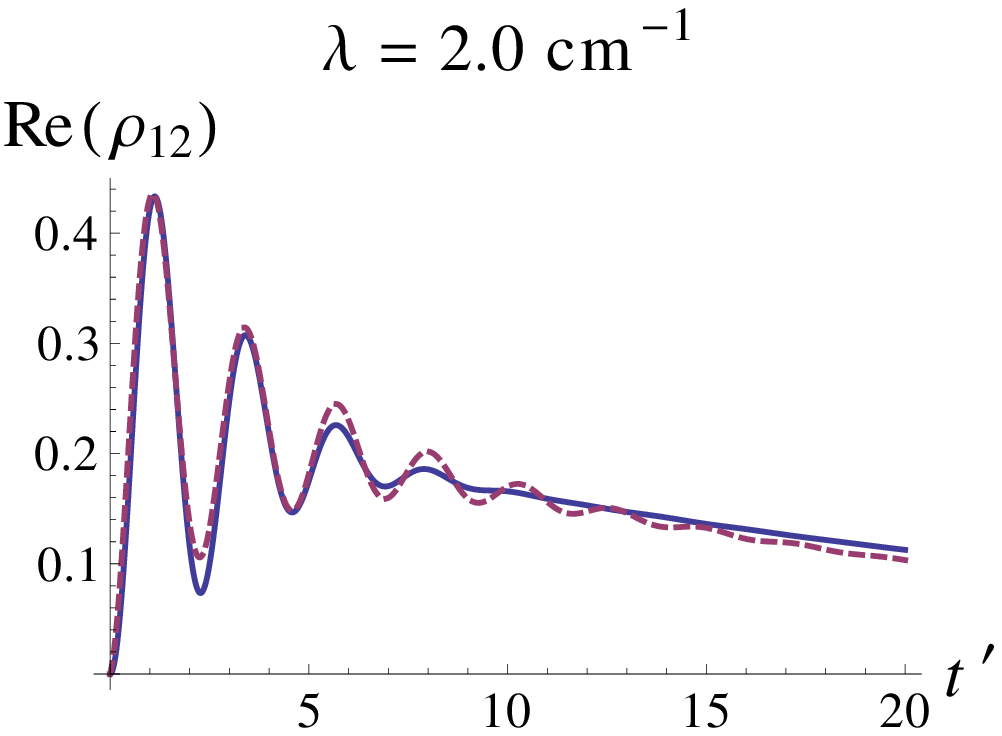}&
\includegraphics[height = 3cm, width=  4cm]{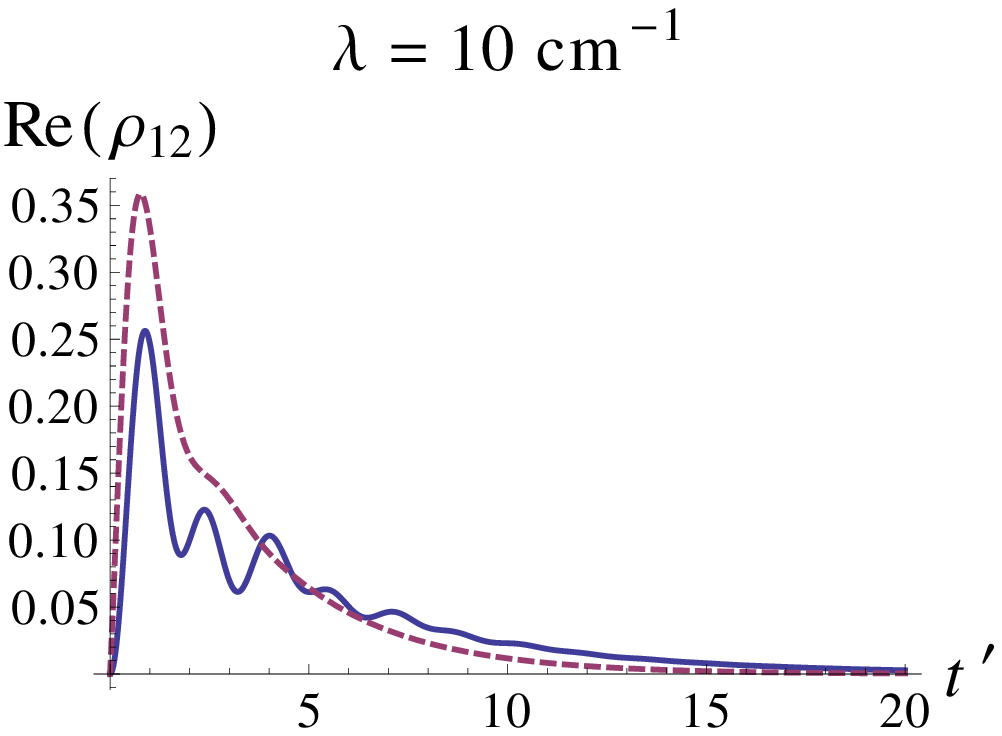}\\
\includegraphics[height = 3cm, width = 4cm]{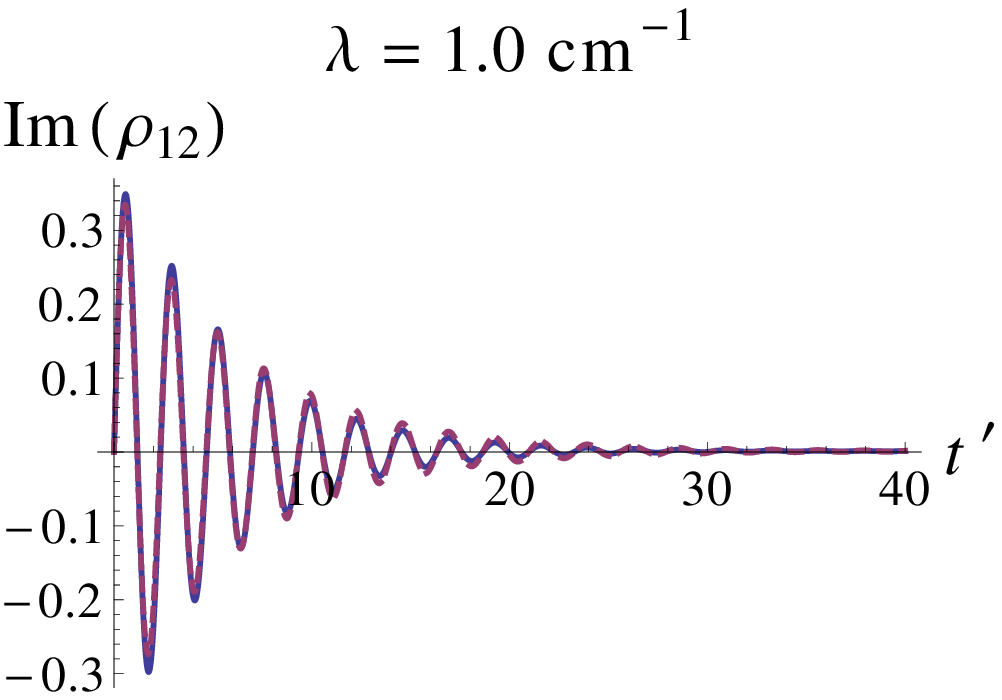}&
\includegraphics[height = 3cm, width = 4cm]{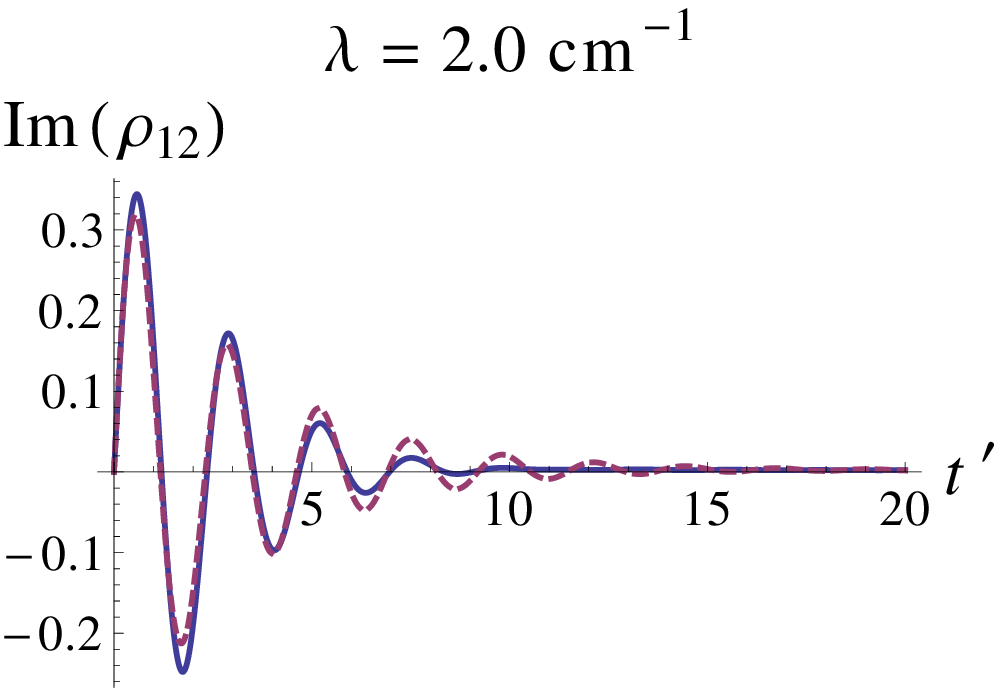}&
\includegraphics[height = 3cm, width = 4cm]{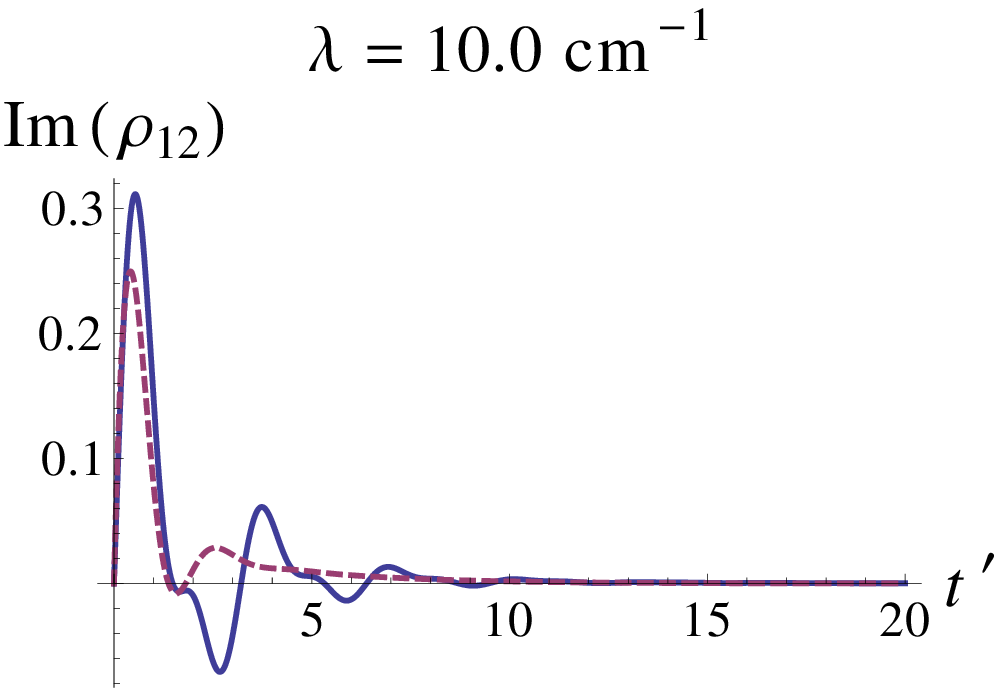}
\end{tabular}
\caption{Time evolution of population on site 1 [$\rho_{11}(t)$]
and the coherences [$\rho_{12}(t)$] [blue (solid) curve is the
non-Markovian solution and red (dotted) curve is the Markovian
approximation], for various values of $\lambda$ (in cm$^{-1}$). Other parameter are: $\Delta$ = 100 cm$^{-1}$,
$J$ = 50 cm$^{-1}$,~~$\gamma = 10^{13}$ s$^{-1}$.}\label{det1}
\end{figure}

\begin{figure}[htbp]
\centering
\begin{tabular}{ccc}
\includegraphics[height = 3cm, width = 4cm]{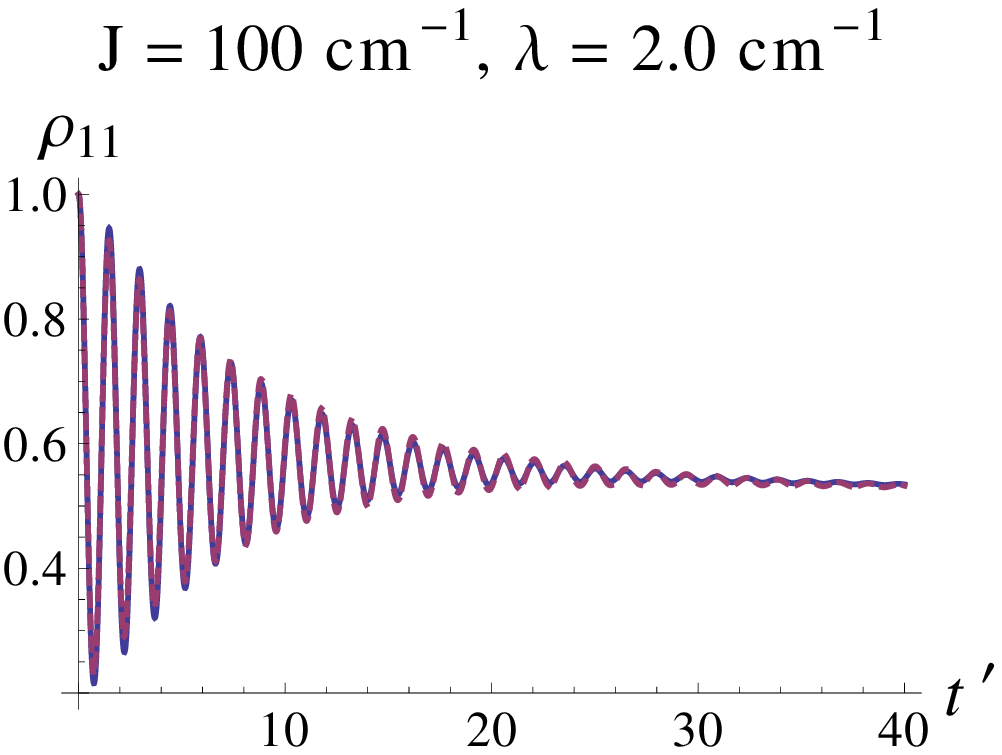}&
\includegraphics[height = 3cm, width = 4cm]{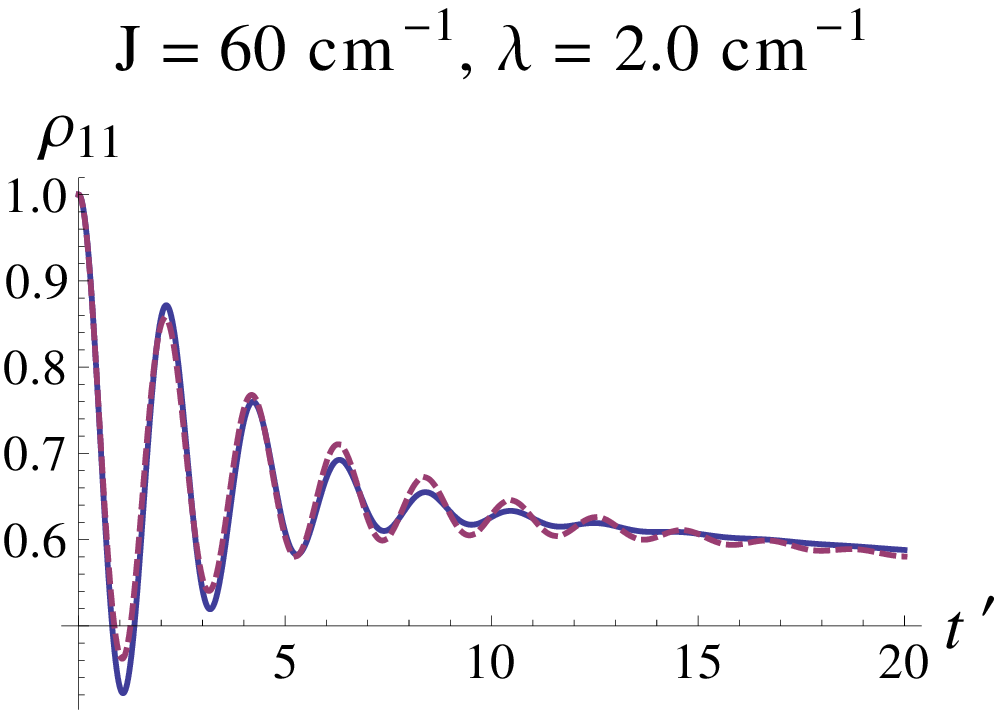}&
\includegraphics[height = 3cm, width = 4cm]{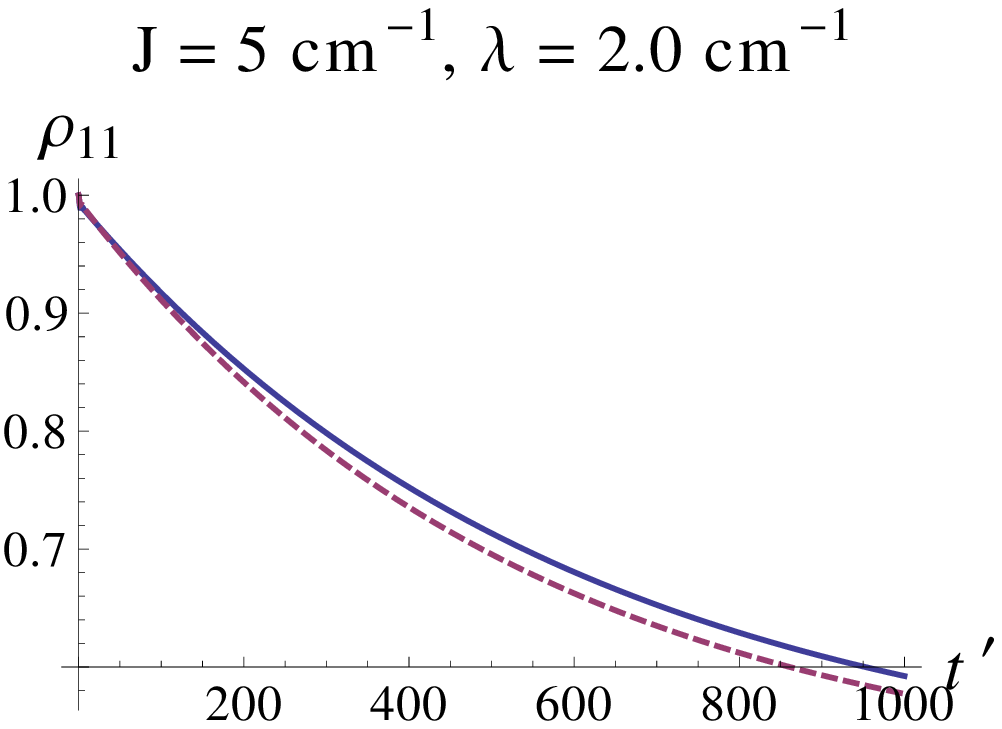}\\
\includegraphics[height = 3cm, width = 4cm]{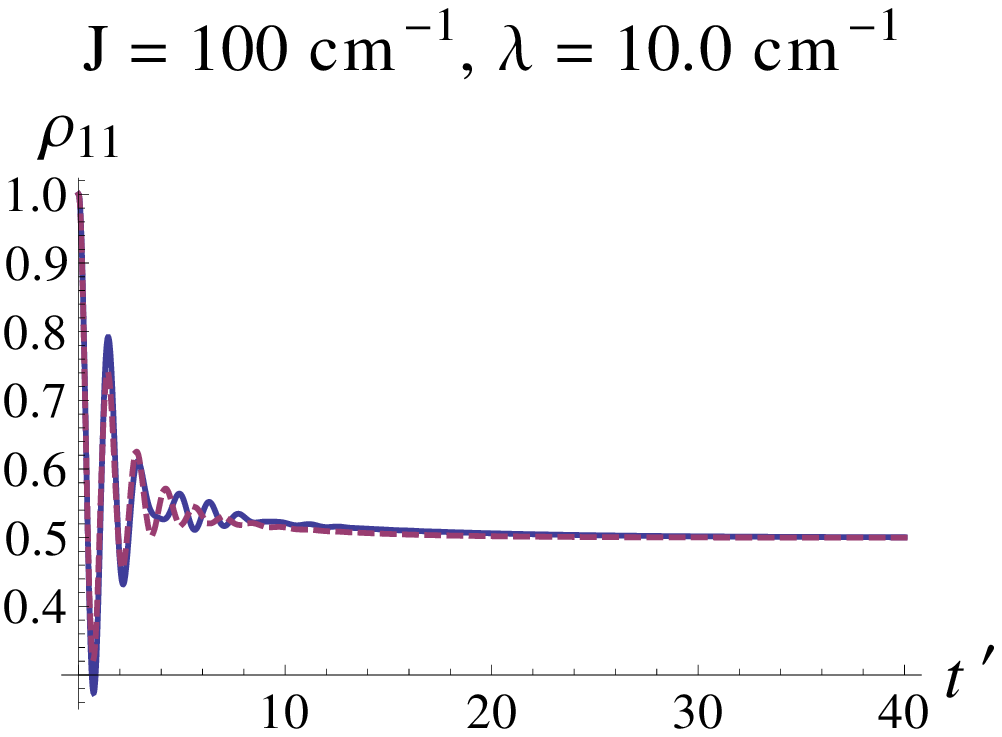}&
\includegraphics[height = 3cm, width = 4cm]{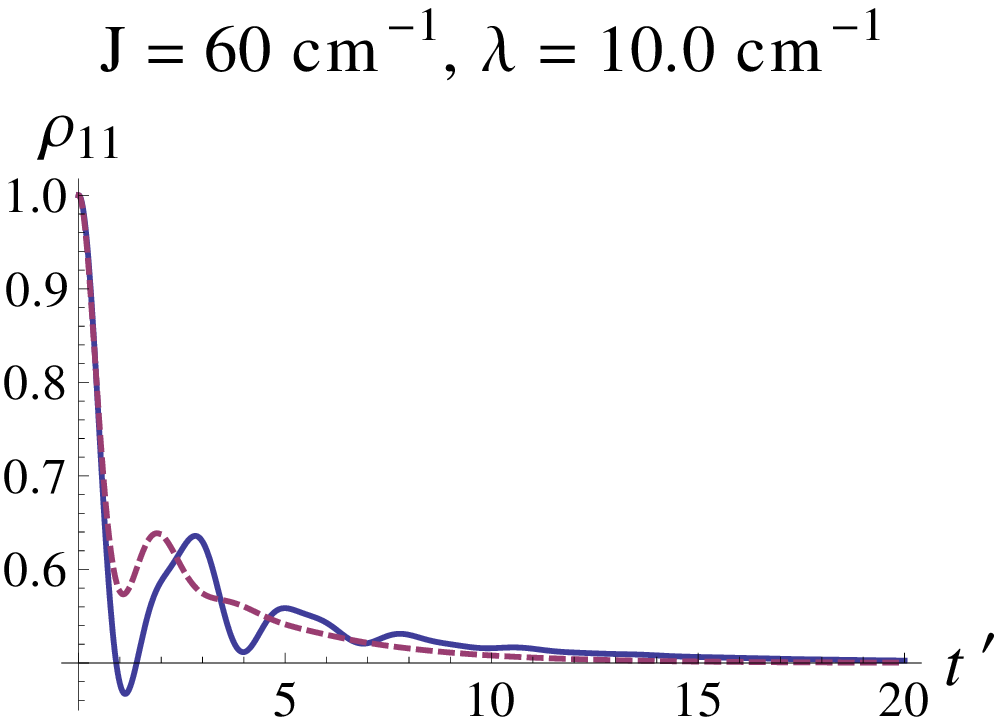}&
\includegraphics[height = 3cm, width = 4cm]{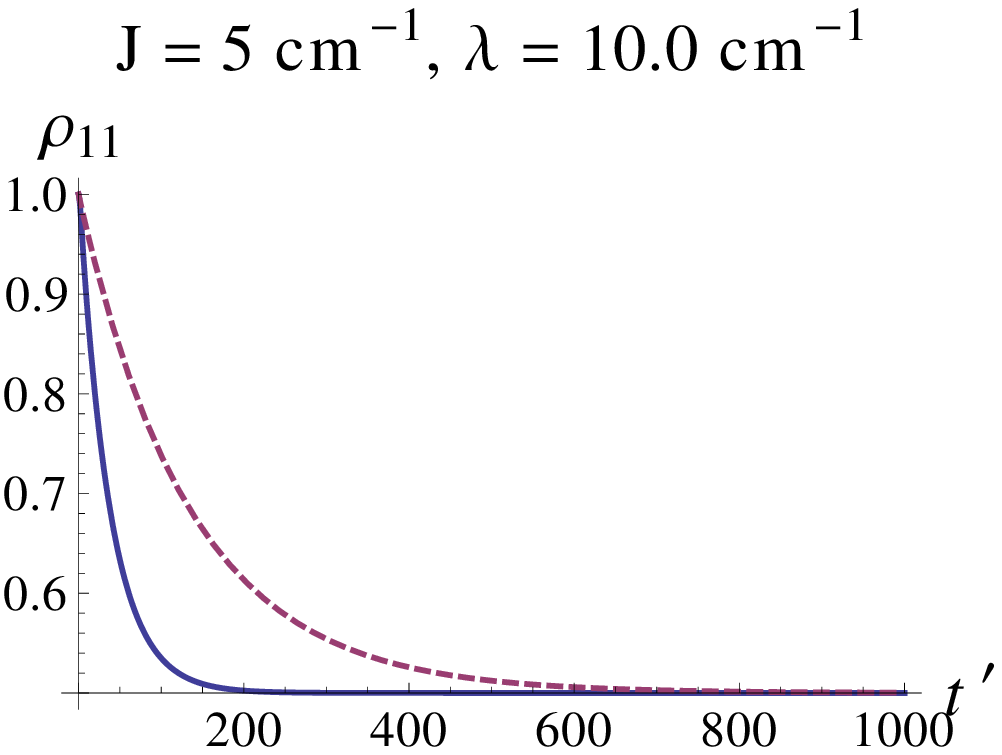}
\end{tabular}
\caption{Time evolution of population on site 1: blue (solid)
curve is the non-Markovian and red (dotted) curve is the
Markovian) for various values of the reorganization
energy $\lambda$ and inter-site coupling $J$.} \label{det2}
\end{figure}

\begin{figure}[htbp]
\centering
\begin{tabular}{ccc}
\includegraphics[height = 3cm, width = 4cm]{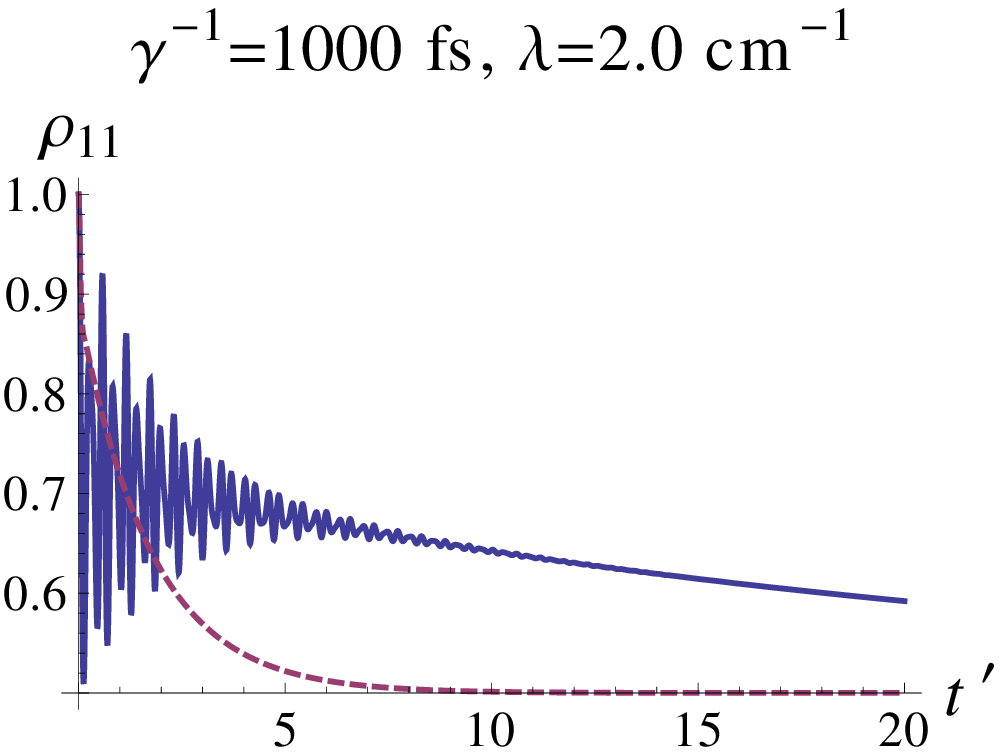}&
\includegraphics[height = 3cm, width = 4cm]{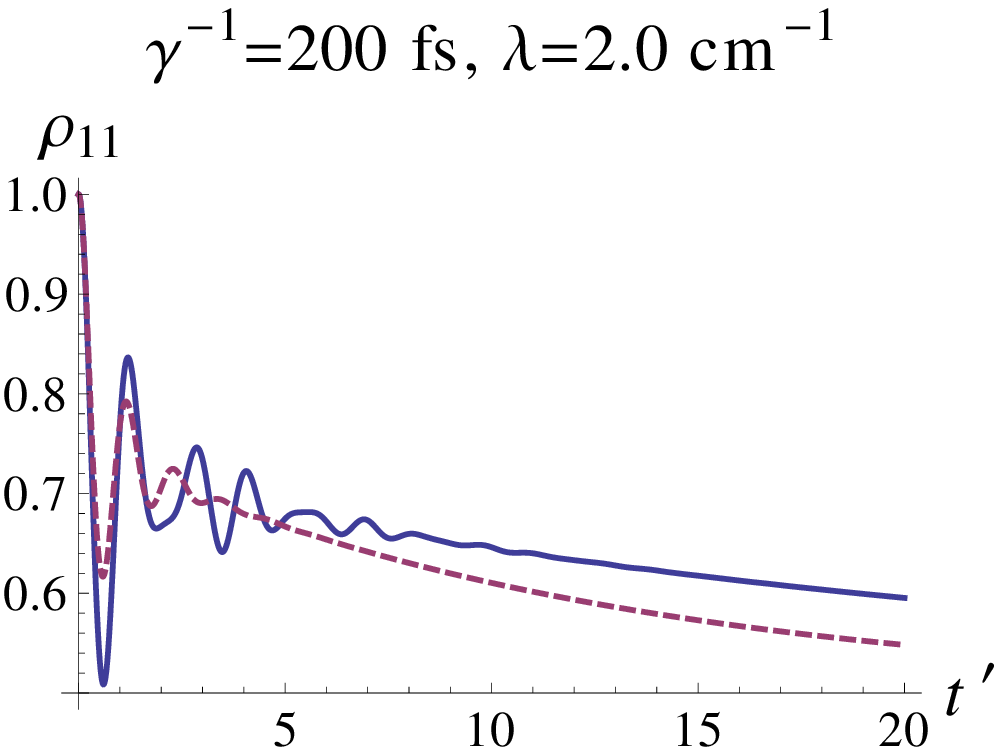}&
\includegraphics[height = 3cm, width = 4cm]{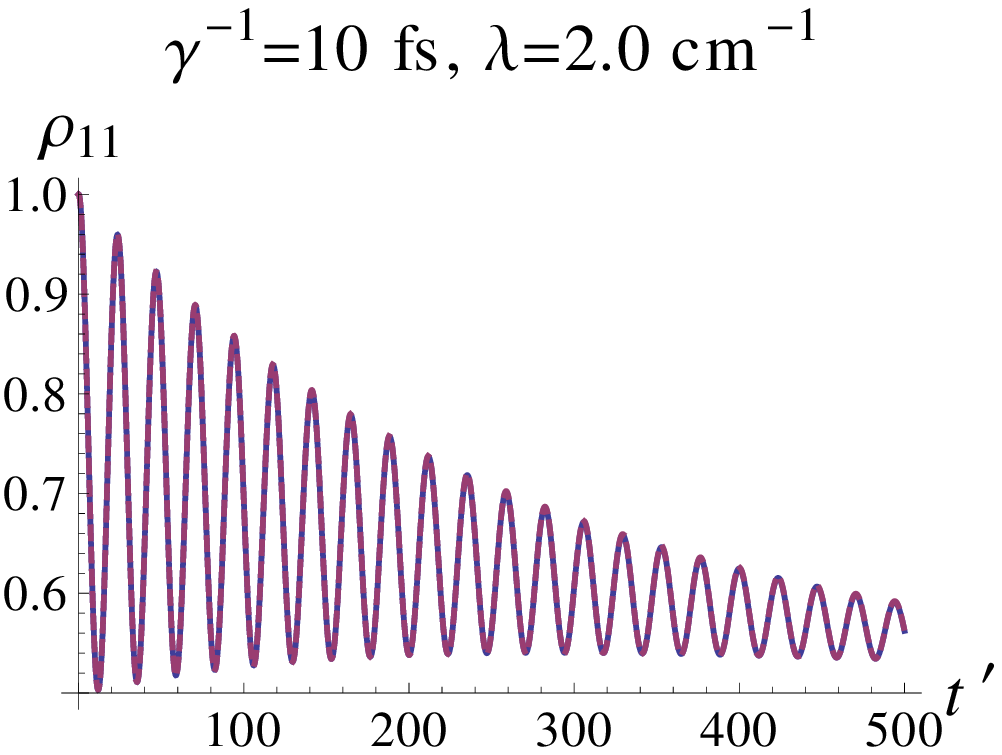}\\
\includegraphics[height = 3cm, width = 4cm]{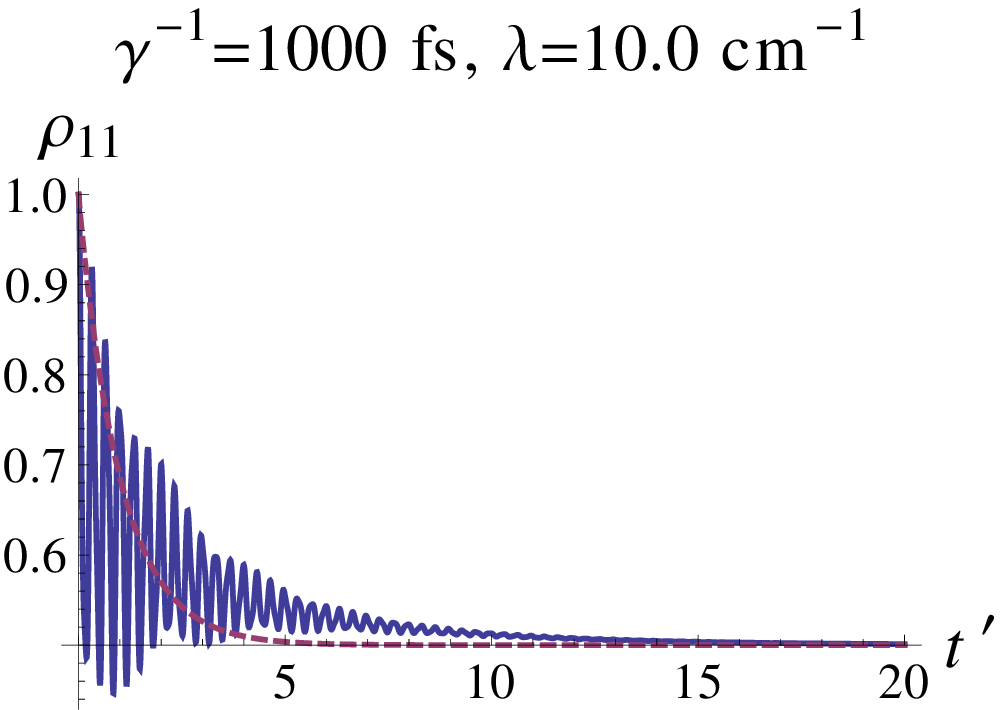}&
\includegraphics[height = 3cm, width = 4cm]{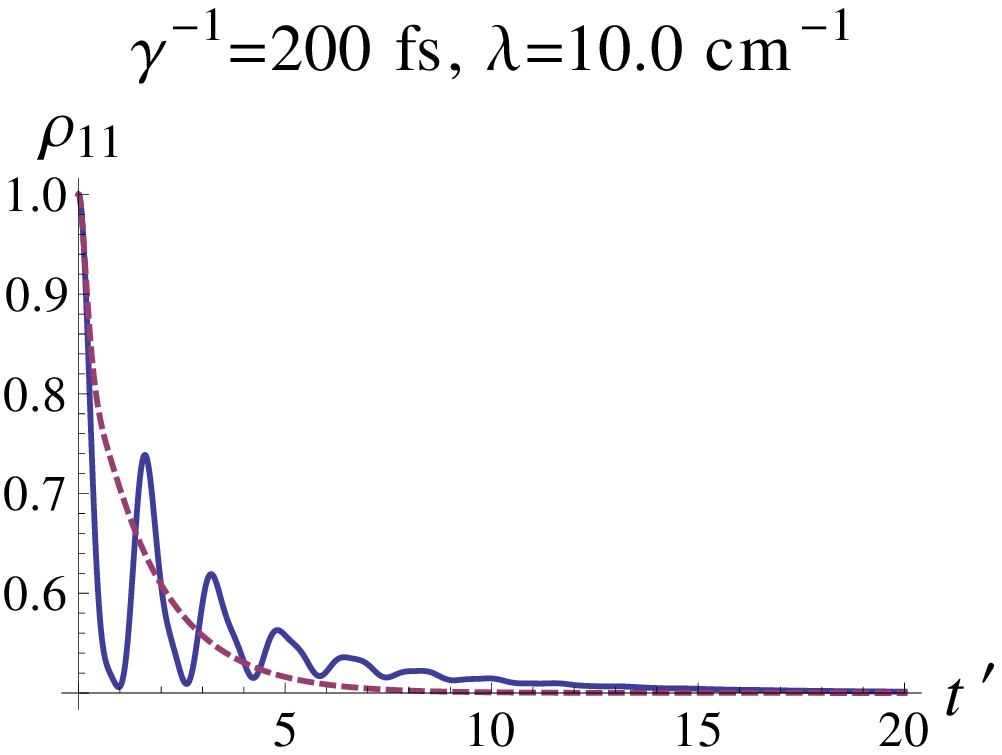}&
\includegraphics[height = 3cm, width = 4cm]{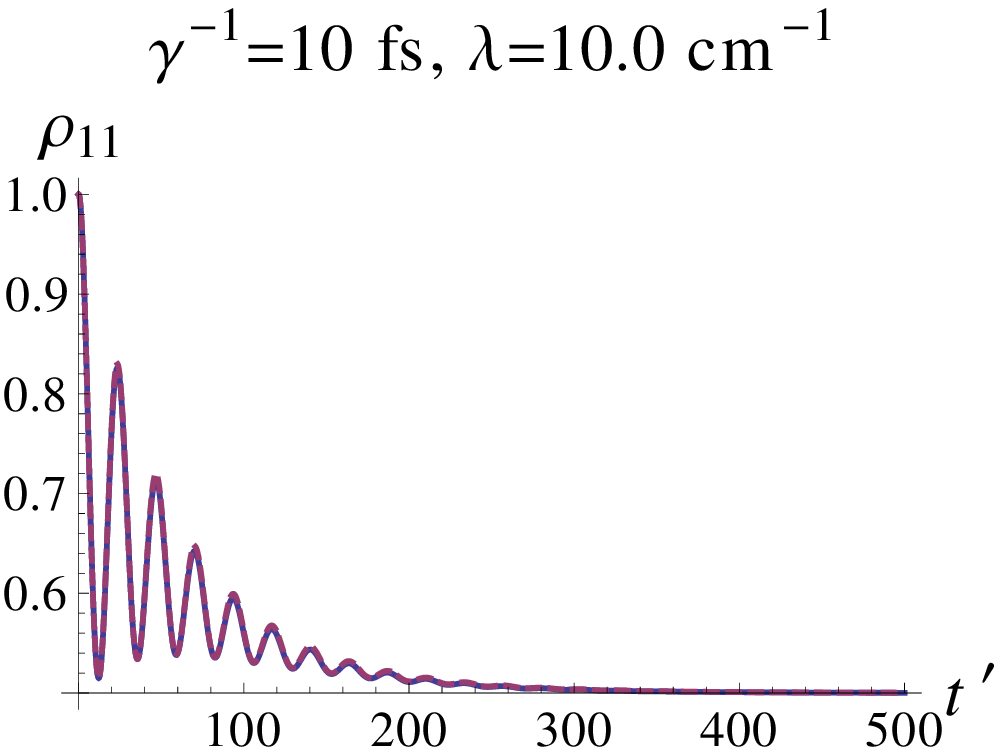}
\end{tabular}
\caption{Time evolution of population on site 1 [blue (solid)
curve is the non-Markovian and red (dotted) curve is the
Markovian] for various values of the reorganization
energy $\lambda$ and phonon relaxation time $\gamma^{-1}$.}\label{det3}
\end{figure}

To investigated the validity regime of Markovian approximation beyond the above two limiting cases we have to
rely on the numerical computation (as the analytic approach becomes very cumbersome). We here display an
extensive list of graphs (Figs.~\ref{det1},\ref{det2}, and \ref{det3}) which explore how Markovian approximation
behaves for various values of parameters.

%------------------------------------fig.--------------------------

%------------------------------------fig. --------------------------

%------------------------------------fig. --------------------------

First consider the non-Markovian results (solid curves). Coherent oscillatory
dynamics up to substantial time scales are evident. Oscillatory behavior in the
populations are seen (Fig. \ref{det1}) to be accompanied by oscillations in the
off diagonal elements $\rho_{12}$ representing coherences. The oscillations fall off faster with
increasing re-organization energies, as expected. Another point to be noted is that the relaxation to
equilibrium populations occurs on a longer time scale than does relaxation of the coherences to
zero. This difference is more evident at smaller values of
re-organization energy $\lambda$. The dependence of population relaxation on
the bath correlation decay time $\gamma^{-1}$ is shown in Fig. \ref{det3}.
Clearly, the larger the $\gamma$ (i.e., fast bath relaxation), the better is the Markov approximation.

Figures \ref{det1} - \ref{det3} also contain a comparison of the Markovian limit
to the non-Markovian solution for domains other than those in Table I.
For the standard electronic coupling parameter values in photosynthetic EET:
$\gamma^{-1} = 100~\textrm{fs},~~J = 50 ~\textrm{cm}^{-1},~~
\Delta = 100~\textrm{cm}^{-1},~~T =300~$K, figure \ref{det1} shows that
the Markovian approx is very good for $\lambda = 1 ~ \textrm{cm}^{-1}$,
fair for $\lambda = 2 ~ \textrm{cm}^{-1}$, and
invalid for reorganization energies $\lambda \geq 10 ~ \textrm{cm}^{-1}$. As typical values of
$\lambda$ in photosynthetic EET systems are considerably larger than $\lambda =
10$ cm$^{-1}$, these results support the conclusions of Ref. \cite{if}, 
although with a different approach.

Figs. \ref{det2} and \ref{det3} show the validity of the Markov approximation obtained by varying $(J,\lambda)$
and $(\gamma^{-1}, \lambda)$ respectively, but keeping $\Delta=100$ cm$^{-1}$. The results show
that the Markovian approximation is poor for large $\lambda$ and small $J$, and for large $\gamma^{-1}$ and
small $\lambda$. Other parameter values can be easily examined using this approach.

\begin{figure}
\includegraphics[height = 5cm, width = 13cm]{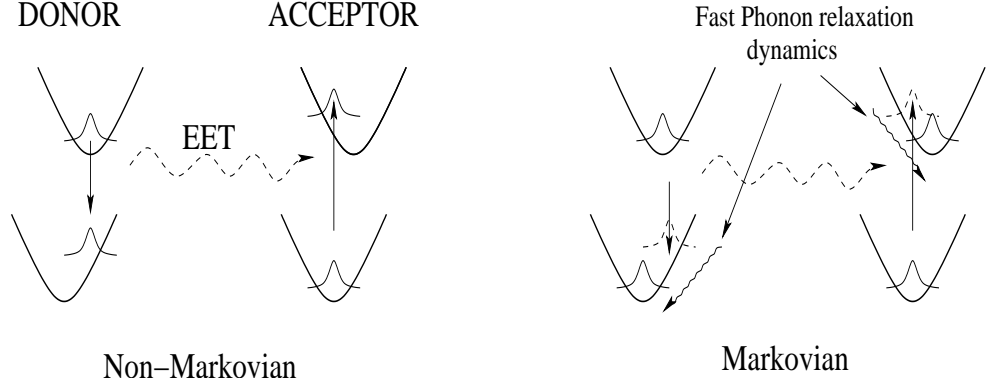}
\caption{Approximate Physical Picture}
\label{phy}
\end{figure}

From these qualitative conclusions a rough physical picture can be drawn as depicted in Fig.~(\ref{phy}). In
the Markovian regime, after photo-excitation, bath (nuclear co-ordinates) re-organize very fast representing an
''apt" bath (right hand picture) while in the non-Markovian regime bath correlations exist for longer time scale
(of the order of exciton transfer time scale). Thus the combined (system+bath) dynamics is much more complicated
in the non-Markovian regime.

%------------------------------------fig.--------------------------
%---------------------------------------

Here, results are given for the particular initial conditions:  $ \rho_{11}(0) = \td{x}(0) =1,
~~\td{y}_1(0) =  \td{y}_2(0)=0,
~~f_1(0)=f_2(0)=f_3(0)=\dot{f}_1(0)\dot{f}_3(0)=0$.
These initial conditions (corresponding to
all the population being on site 1, and no
coherences) are those which have been used extensively in
literature\cite{if} but are somewhat unphysical\cite{if3}, because they lack initial coherences which become
important in preparatory photo-excitation. We have considered this problem in the second reference of
the list\cite{nav} which considers photo-excitation of a dimer oscillator system with an ultra-short laser
pulse. It appeared that the presence of initial coherence ( at $t' = 0$) effected the time scale on which the
populations reach equilibrium value but had little effect on the overall damping-out of the coherences (see for
details\cite{nav}).

\section{Phenomenological approach: a stochastic model}

In the previous sections we saw that 2nd Born quantum master equation cannot be applied to the real light
harvesting systems as in these systems the system-bath coupling ($\lambda \sim 100~cm^{-1}$) is of the same
order of magnitude of the system-system coupling ($J\sim 100~cm^{-1}$). To make the situation more intractable,
it is not possible to justify the Markovian approximation when $\lambda \gg 1~cm^{-1}$ (given that Markovian
master equations are much easier to solve than the non-Markovian ones). Thus the use of Markovian Redfield
theory to these systems is questionable as pointed out in\cite{if,nav}. This open up a difficult problem. One
should formulate some non-perturbative theories.  Recently Ishizaki and Fleming\cite{if} have developed a
formalism which goes beyond the limitations of the 2nd Born master equation. They use the
reduced hierarchy equation approach previously developed by 
Tanimura and Kubo\cite{tk}. There is an other route to the problem pioneered by people like Silbey\cite{silbey}.
In this approach one uses a unitary transformation (called polaron transformation) to completely eliminate the
system-bath coupling Hamiltonian. But this re-normalize the system Hamiltonian. Then one re-partition the
resulting system Hamiltonian to identify a weaker term which can be used as a perturbation. The remaining
problem is done in line with 2nd Born master equation\cite{jang}.

We have developed an alternative stochastic approach which is phenomenological in nature\cite{nav3}. This
approach, as its input, takes the homogeneous line width from the experiment and uses Kubo's stochastic theory
of motional narrowing to get phenomenological decoherence rate.

\subsection{The model and its solution}
\begin{figure}
\includegraphics[height = 4 cm, width = 13cm]{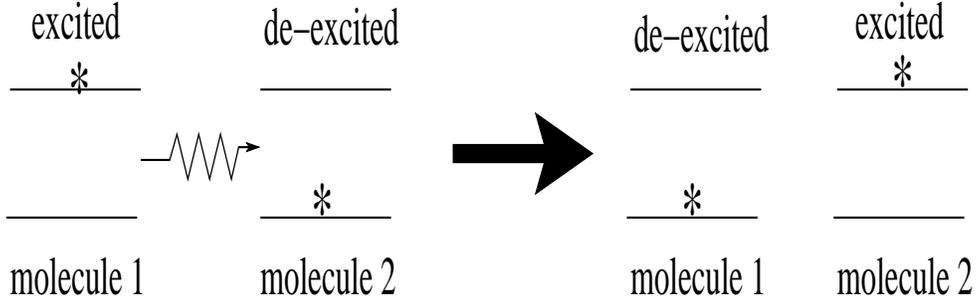}
\caption{Two interacting molecules}\label{mole}
\end{figure}

We again consider the dynamics of exciton transfer between two molecules modeled as two-level electronic
systems (Fig.~\ref{mole}). These two-level systems are electronically coupled with each other with coupling
$J$. 

Due to the electronic coupling between the molecules the exciton will transfer back-and-forth between the
molecules. This will happen forever if the molecules are completely isolated---pure oscillatory quantum motion. 
Now consider that our two-molecular system is open i.e., interacting with the external bath 
degrees-of-freedom----the phonons. It is well known that the dynamics remain quantum at short time scales and
becomes classical at longer time scales\cite{rush}. This quantum-to-classical crossover happens at a critical
time scale which is inversely proportional to system-bath coupling energy ($t_c \propto \frac{\hbar}{\eta}$).
Large $\eta$ (system-bath coupling) means fast quantum-to-classical crossover and vice versa.  We denote
system-bath coupling strength with $\eta$ in the subsequent subsections (before we used $\lambda$).

In a recent contribution\cite{nav3} we extracted $\eta$ from experimental information using Kubo's stochastic
theory of line shapes and observed upto what timescale one could see quantum effects.

The important point is that we modeled the dynamical effect of phonons as a stochastic noise. The total
Hamiltonian takes the form 
\begin{equation}
H = \ep_1 |1\ra\la 1| + (\ep_2+\ep(t)) |2\ra\la2| + J (|1\ra\la2|+|2\ra\la1|).
\label{eq1}
\end{equation}
Here $\ep_1$ is the energy of the upper electronic level of the first molecule and $\ep_2$ is that of the
second molecule. The ground state energies of both the molecules are taken to be zero. The energy separation
$\ep_2-\ep_1$ has a random component (due to phonons) which we denote with $\ep(t)$. $\ep(t)$ is a stochastic
process taken here as Gaussian White Noise (GWN): 
\begin{equation}
\la \ep(t) \ra = 0, ~~~ \la \ep(t) \ep(\tau) \ra = \hbar^2\eta \delta(t-\tau).
\end{equation}
Here, as mentioned before, $\eta$ is the strength of system-bath coupling (also known as dynamical disorder)
measured in the units
of frequency.

We start with Liouville-von-Neumann equation for the total density matrix,
\begin{equation}
i\hbar\frac{\pr \hat{\rho}(t)}{\pr t} = [\hat{H},\hat{\rho}(t)].
\end{equation}

As $\hat{H}$ is a stochastic operator, thus $\hat{\rho}(t)$ is also a stochastic operator. Therefore, 
we need to evaluate the averaged density matrix. So we need to do an averaging over the dynamical disorder
which is denoted by $ \la ...\ra $. We define  $\varsigma(t) \equiv \la \rho(t)\ra$.

Averaging over dynamical disorder:
\begin{equation}
i\hbar \frac{d\varsigma(t)}{dt} = \underbrace{\la H \rho(t)\ra}_{Term I} - \underbrace{\la \rho(t) H\ra}_{Term
II}.
\end{equation}

In term I and II above we have terms like $\la \rho(t)\ep(t)\ra $. As $\rho(t)$ is a functional of
$\ep(t)$ (a stochastic quantity) the $\rho(t)$ will also be a stochastic function. To decouple these we will use
the famous theorem of Novikov\cite{novi}:

\begin{equation}
\la \ep(t) \rho_{ab}(t)\ra = \int_{-\infty}^{\infty} dt' \la \ep(t) \ep(t')\ra \left\la \frac{\delta
\rho_{ab}(t)}
{\delta \ep(t')}\right\ra
\end{equation}
Here $\frac{\delta \rho_{ab}(t)}{\delta \ep(t')}$ is the functional derivative. Using the  properties of
stochastic noise and with some simplification(see appendix A), we get

\begin{equation}
 \la \ep(t)\rho_{12}(t)\ra = i\hbar \eta \varsigma_{12}(t).
\end{equation}

Finally, one has a set of coupled differential equations:
\begin{eqnarray}
&&\frac{d\varsigma_{11}(t)}{dt} = -i(J/\hbar) (\varsigma_{21}(t)-\varsigma_{12}(t)) \nonumber\\
&&\frac{d\varsigma_{12}(t)}{dt} = -i(\Delta/\hbar) \varsigma_{12}(t) - i (J/\hbar)
(\varsigma_{22}(t)-\varsigma_{11}(t))-\eta \varsigma_{12}(t) \nonumber\\
&&\varsigma_{11}(t)+\varsigma_{22}(t)=1.
\end{eqnarray}
The above system of ODEs can be solved analytically, however, the exact 
expression is very cumbersome. We give the analytic solution only in the long time limit (see
Appendix B).

To simulate the dynamics of decoherence in this dimer model with physical parameters of the FMO
problem ($
\Delta =\ep_1-\ep_2 \simeq 100 cm^{-1},~J\simeq 100 cm^{-1}$), we need to find out our phenomenological
parameter $\eta$, for this we use Kubo's stochastic theory of
lineshapes.

\subsection{Kubo's stochastic theory of lineshapes and estimation of $\eta$ from motionally narrowed lineshape}

\begin{figure}
\begin{tabular}{cc}
\includegraphics[height = 5cm, width = 7cm]{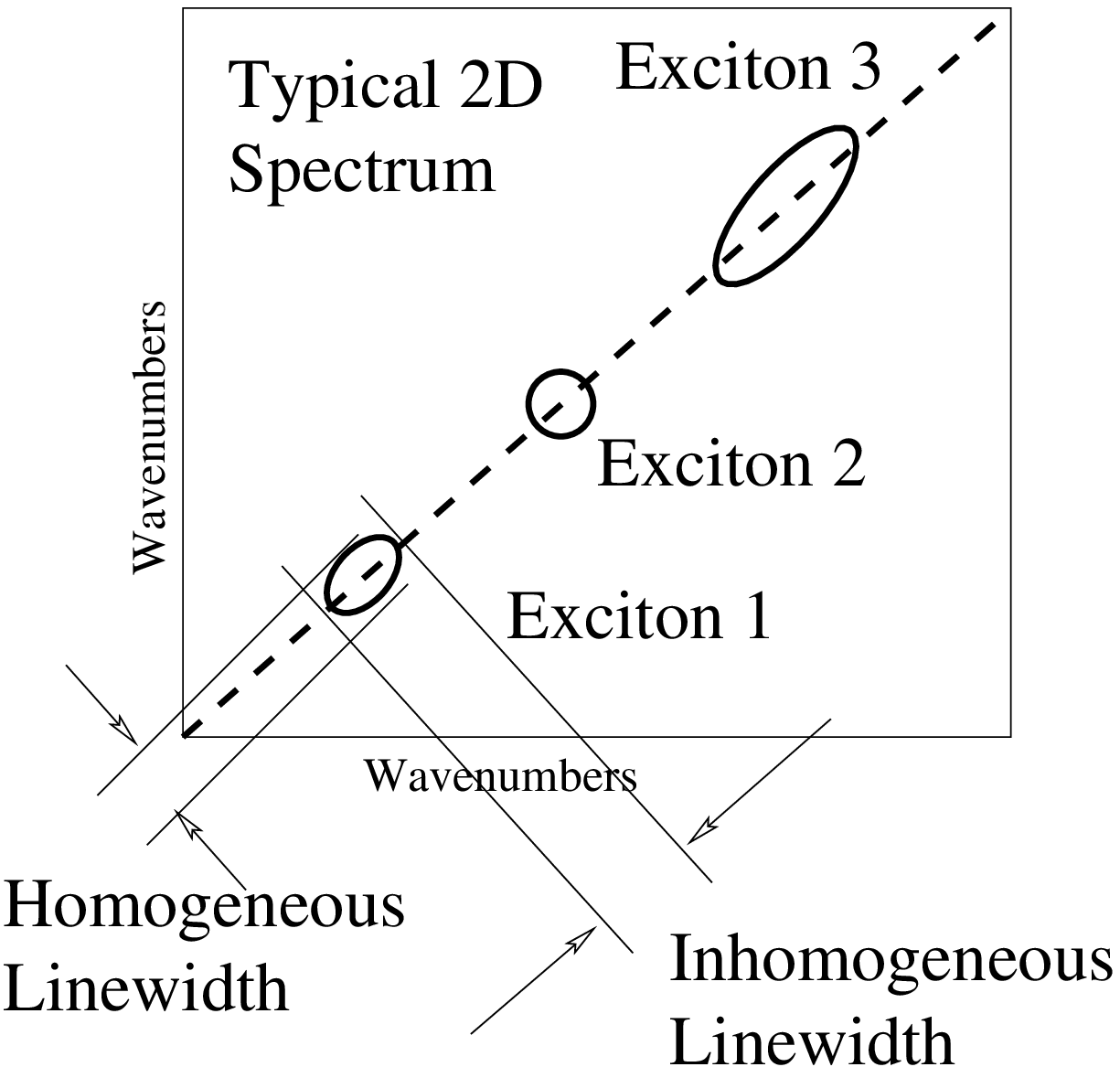}&
\includegraphics[height = 5cm, width = 7cm]{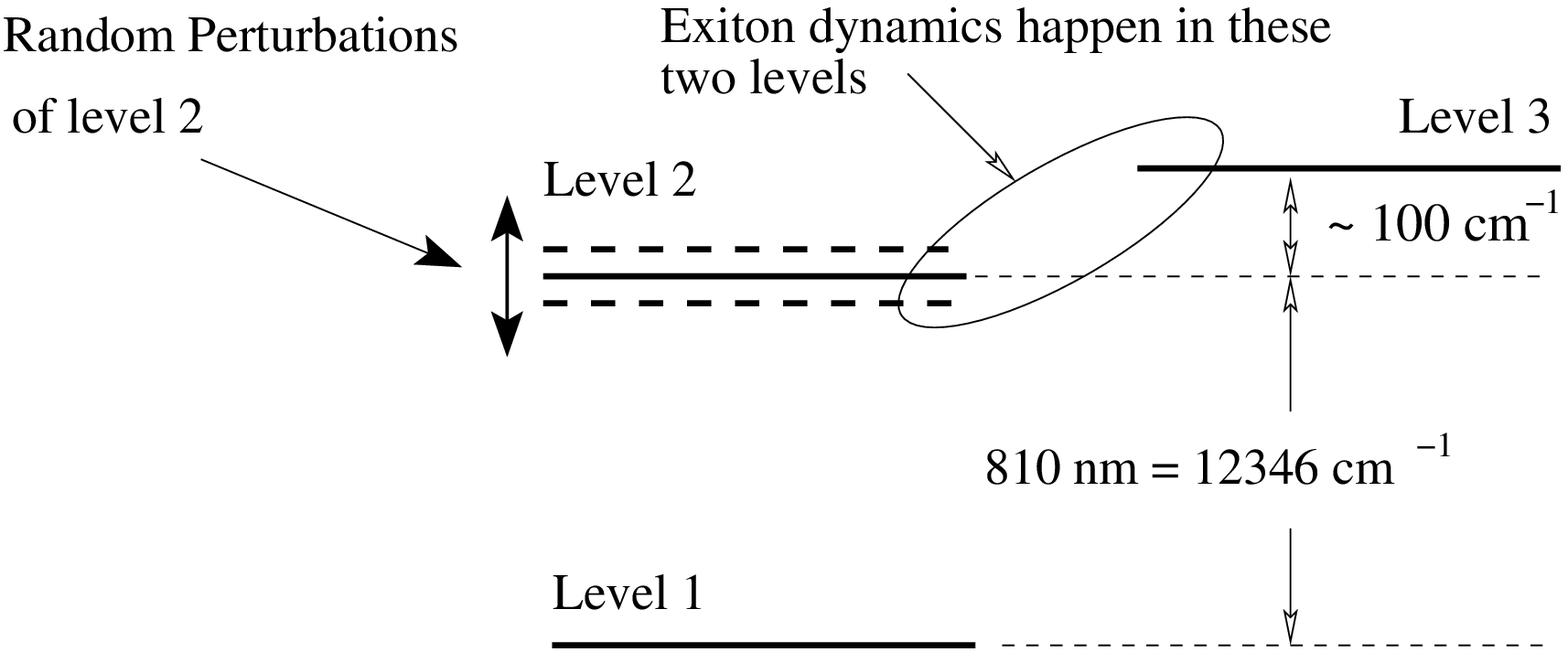}
\end{tabular}
\caption{(Left) Schematic line broadening information in 2-D photon echo spectrum  shown without cross peaks.
The linewidth due to homogeneous and in-homogeneous broadening are in orthogonal directions as shown. (Right)
The energy levels of the model system}
\label{2d}
\end{figure}

We now determine the phenomenological parameter $\eta$. Let us focus on exciton 1 in 2-D photon echo spectra
which occurs at $810 ~nm$ (Fig.~\ref{2d}). 

We use Kubo's randomly modulated oscillator model for the exciton under question. 
The energy levels of the model system are given in Fig.~\ref{2d}. The levels 1 and 2 are separated {\it on
the average} by $\om_0 = 810 ~nm$.
% 
% \begin{figure}[h!]
% \includegraphics[height = 5cm, width = 9cm]{f3.eps}
% \label{levels}
% \end{figure}

Let the level 2 be randomly modulated with a random process $\om_1(t)$ such that  $\lim_{T\rta \infty}
\frac{1}{T}\int_0^T \om_1(t) dt = 0$. The random frequency of the level 
2 is then given as $\om(t) = \om_0 + \om_1(t)$. We use the classical oscillator model for the molecule 
with resonance frequency $\om_0 = 810 ~nm$. The equation of 
motion is  $\dot{x}(t) = i \om(t) x(t)$,  with  solution 
\begin{equation}
x(t) = x(0) \exp{\left( i \om_0 t + i \int_0^t \om_1(t')dt' \right)}.
\end{equation}
The absorption spectrum measures a large number of possible realizations of the random process $\om_1(t)$ 
( The radiation field acts on a  macroscopic number 
of molecules present with different realizations of $\omega_1(t)$ in the sample). Thus one need to
consider an ensemble average to compare with absorption spectrum:
\begin{equation}
\la x(t)\ra = x(0) e^{i \om_0 t} \left\la \exp{ \left(i \int_0^t \om_1(t')dt' \right)}\right\ra,
\end{equation}
and the average time correlation is given by $\la x(t) x^\ast(0)\ra = |x(0)|^2 e^{i \om_0 t} \phi(t)$ where
\begin{equation}
\phi(t) = \left\la  \exp{ \left(i \int_0^t \om_1(t')dt'\right)}\right \ra 
\end{equation}
is called the relaxation function of the oscillator. The absorption spectrum is
\begin{equation}
I(\om-\om_0) = \frac{1}{2\pi}\int_{-\infty}^{+\infty} e^{- i (\om - \om_0)t} \phi(t) dt. 
\end{equation}
This is the direct consequence of the famous fluctuation-dissipation theorem. The temporal  character of the
decay of 
fluctuations tells directly the dissipative characteristics  
of the system. The intensity distribution $I(\om)$ will be broadened by the stochastic process $\om_1(t)$. 
To define the stochastic process $\om_1(t)$ let $P(\om_1) d\om_1$ be
 the probability to find the random frequency $\om_1$ to be in the range $\om_1$ to $\om_1 + d\om_1$ when 
picked randomly from the ensemble.  In Kubo's theory, the stochastic process is 
defined with two parameters (1) magnitude of modulation $\Delta^2 = \int \om_1^2 P(\om_1)d\om_1 = \la
\om_1^2\ra$ 
and (2) correlation time $\tau_c = \int_0^\infty f_c(t)dt$
 where correlation function is defined as $f_c(\tau) = \frac{1}{\Delta^2} \la\om_1(t)\om_1(t + \tau) \ra$. 

It is well known that for a Gaussian process, relaxation function can be written in terms of the correlation
function:

\begin{equation}
 \phi(t) = \exp{\left(- \Delta^2 \int_0^t (t -\tau) f_c(\tau) d\tau\right)}
\end{equation}
In the present case we have considered Gaussian White Noise (GWN) which has zero correlation time. 
 Our case corresponds to the fast modulation case of Kubo  $\tau_c <<\frac{1}{\Delta}$\cite{sto}. The
correlation 
function decays very fast and the upper limit of the integral
 in the above equation can be extended to $\infty$. 
This leads to $\phi(t) \propto e^{- \Delta^2 \tau_c |t|}$. This results in the famous narrowing of the
lineshape 
from the Gaussian to Lorentzian form. In the present case we have $\la
\frac{\ep(t)}{\hbar} \frac{\ep(t +\tau)}{\hbar}\ra 
= \eta \delta(\tau) $ which leads to $\phi(t) = 
e^{-\eta |t|}$. Comparison with the previous $\phi(t)$ shows that $\eta = \Delta^2 \tau_c$ which gives the 
rate of decay of the correlation function. In our GWN case 
$\Delta \rta \infty$  since white noise contains all frequencies and $\tau_c \rta 0$ (delta correlated noise) but 
$\Delta^2 \tau_c$ is finite and is equal to $\eta$.

Thus the absorption spectrum takes the form

\begin{equation}
I(\om-\om_0) = \frac{1}{\pi} \frac{\eta}{(\om-\om_0)^2 + \eta^2}.
\end{equation}

This is the famous Lorentzian lineshape narrowed from the Gaussian shape (called motional narrowing).

Our aim is to find out our phenomenological parameter $\eta$. Thus we need to fit this 
$I(\omega-\om_0)$ with the real experimental observation  and to extract $\eta$. We will use this to simulate
the quantum dynamics of the density matrix elements. 

The basic problem with linear absorption line shape is that it is broadened both by homogeneous
and in-homogeneous mechanisms. {\it In our case the 
broadening is homogeneous due to dynamical disorder} and thus we need to subtract the in-homogeneous component  
due to static disorder.  But thanks to the 2-D photon echo spectroscopy one has the important information about
both homogeneous and in-homogeneous broadening (see Fig.~\ref{2d}). A brief introduction to the physics of 2D
photon echo spectroscopy is given in the appendix c (for details see for example\cite{2dphoton}). We want to
measure Full Width at Half Maximum (FWHM) of the homogeneously broadened  peak\cite{note1}. We consider Fig 2
(a) of G. S. Engel et. al\cite{gs}. The homogeneous broadening is along the main diagonal (see Fig.~\ref{2d}).
From the scale given in terms of nano-meters of the figure 2(a) of G. S. Engel et. al., the FWHM is about
$\simeq 10 nm$ 
and the exciton peak occurs at $810 ~nm$. This gives the frequency broadening $\delta \om_{FWHM} \simeq 2.87 
\times 10^{13} Hz$. 

With this experimental information we plot $I(\om-\om_0)$ such that FWHM is about $\simeq 10 nm = 2.87 10^{13}
Hz$. 
Clearly for the Lorentzian, at FWHM $ \delta \om_{FWHM} = 2 \eta$. This gives $\eta = 0.0143 ~fsec^{-1}$.

\subsection{Long coherences}

We now have all the required  parameters, from the experimental information, namely, 
$\eta = 0.0143 ~fsec^{-1}, ~J = 100 cm^{-1},~{\rm and}~
\Delta= 100 cm^{-1}$. With these values we plot the  dynamics of the density matrix elements $r(t),~ x(t),~{\rm
and}~y(t)$. We clearly see that the density matrix elements show  oscillations upto $500 fsec$, {\it mimicking}
the long coherences observed in the experiments of G. S. Engel et. al. To reproduce the actual spectra observed
for example by G. S. Engel et. al. one has to go beyond this simple two state model. One has to consider a
detailed model of the FMO complex containing not the two coupled molecules (as considered here) but the seven
coupled BChl molecules and there interactions with the protein matrix. This clearly requires a considerable
computational challenge and one has to rely on the numerical approach rather than on an simple analytical
solution as given here.

%------------------------------------fig.SET No.2 --------------------------
\begin{figure}
\centering
\begin{tabular}{cc}
\includegraphics[height = 5cm, width = 7cm]{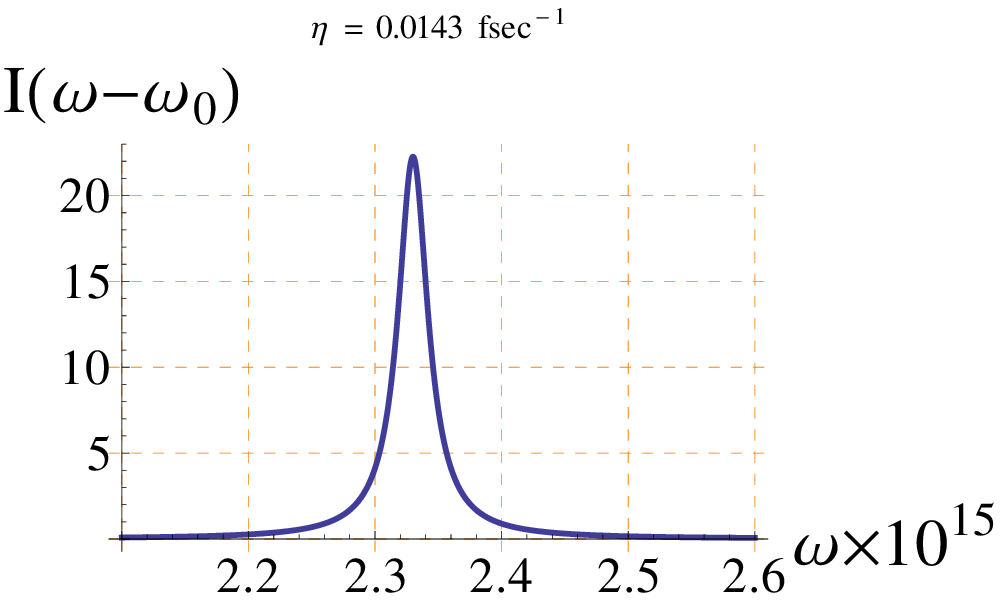}&
\includegraphics[height = 5cm, width = 7cm]{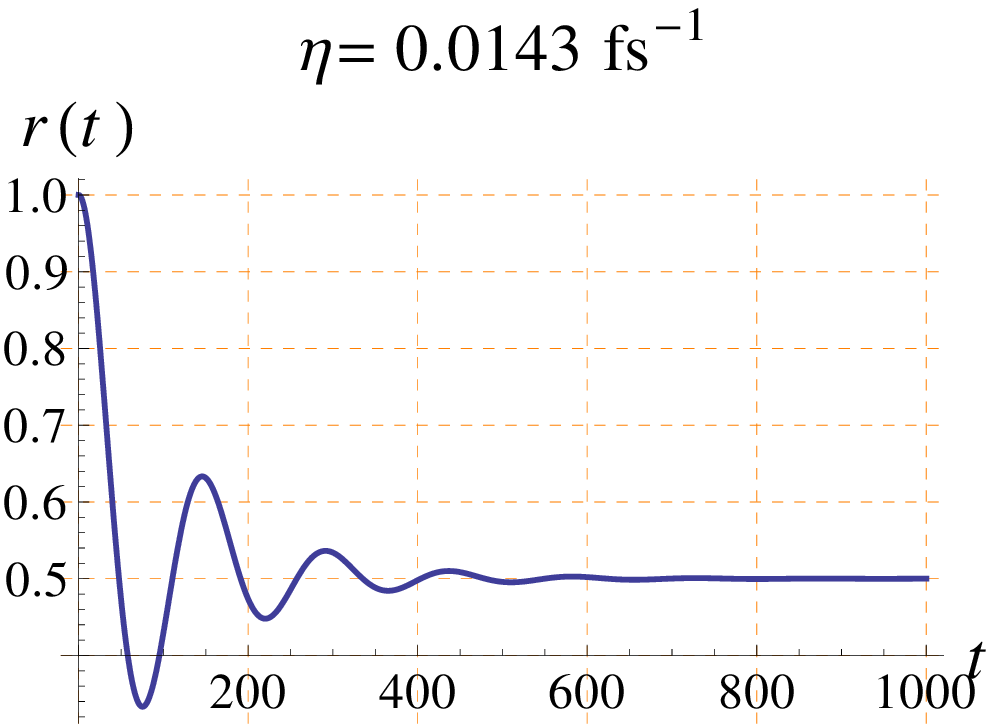}\\
\includegraphics[height = 5cm, width = 7cm]{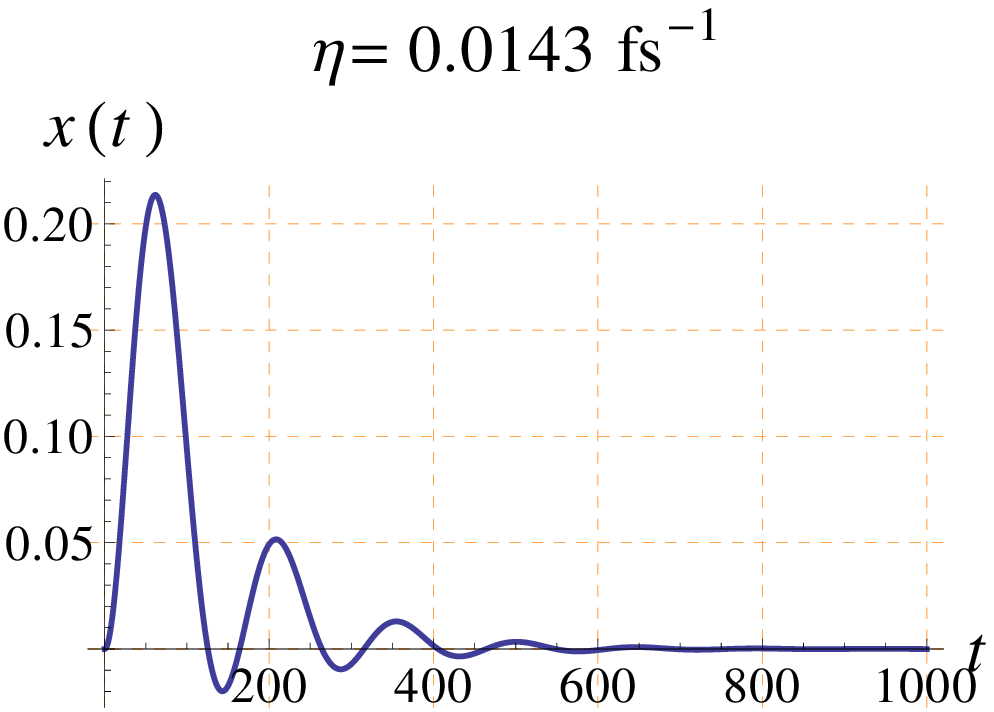}&
\includegraphics[height = 5cm, width = 7cm]{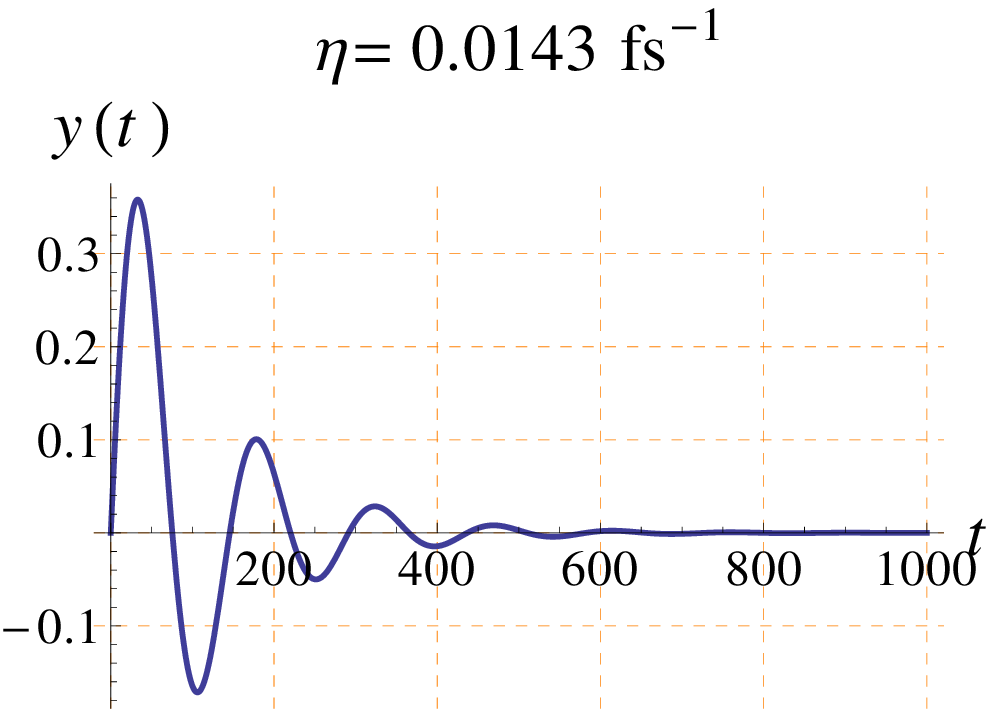}
\end{tabular}
\caption{Line shape function $I(\omega)$ for $\Delta= 100 cm^{-1}, J = 100 cm^{-1},~~\eta =
0.0143~fsec^{-1}$ (upper left). This value of $\eta$ i.e., $0.0143~fsec^{-1}$ give the correct value of
homogeneous line-broadening $\delta \omega \simeq 2.87 \times 10^{13} Hz$.}
\end{figure}
%----------------------------------------------------------------------

\section{Conclusion}

We have seen that 2nd Born quantum master equation fails in modeling the real light harvesting systems. The
reason is that it is perturbative in nature. One has to develop non-perturbative theories for the problem. Also
resent studies with sophisticated 2D photon echo spectroscopy show long livid coherence effects in exciton
motion which contradicts the long held old idea of incoherent motion\cite{dis}. New approaches should explain
this by going beyond the perturbation theories or new mathematical framework is needed.

Our approach towards the 2nd Born master equation enable us (1) to introduce an efficient numerical scheme,
(2) to quantitatively known the validity regime of the Markovian approximation. We saw that it is inadequate
for the present problem. In the last section we introduced a phenomenological approach. In this,  we modeled the
effect of bath as a stochastic noise and its strength was calculated from motionaly narrowed lineshape. For
this we used Kubo's stochastic theory of lineshapes. Significant role was played by the 2D photon echo
spectroscopy as we had the important information about motionaly narrowed lineshape. We saw that the
density matrix elements showed  oscillations upto $\sim 500 fsec$, thus {\it mimicking} the long coherences
effects as observed in recent experiments.

\section{Acknowledgement}
Author would like to thank Prof. Paul Brumer for introducing this problem to him and for many useful
discussions. He is also thankful to Prof. Greg Scholes for pointing out the possibility of motional narrowing
phenomenon in 2D photon echo spectroscopy.

\section{Appendix}
\subsection{Novikov's Theorem}
By taking the matrix elements of equation (19)
\begin{equation}
i\hbar\frac{\pr \rho_{12}(t)}{\pr t} = i\hbar\frac{\pr}{\pr t} \la 1 |\rho(t)|2 \ra = \la 1 | [H,
\rho(t)]|2\ra,
\end{equation}
one can formally solve for $\rho_{12}(t)$ as
\begin{eqnarray}
\rho_{12}(t) - \rho_{12}(0) &=& -\frac{i}{\hbar}\int_0^t ds(\Delta \rho_{12}(s)\nonumber\\
&+& J (1-2\rho_{11}(s))-\ep(s)\rho_{12}(s)).
\end{eqnarray}
Taking the functional derivative $\frac{\delta \rho_{12}(t)}{\delta \ep(s)}$ of the above equation wrt $\ep(t)$
and plugging into
\begin{equation}
\la \ep(t)\rho_{12}(t)\ra = \hbar^2 \eta \int_{-\infty}^{+\infty} ds \delta(t-s) \left\la \frac{\delta
\rho_{12}(t)}{\delta\ep(s)}\right\ra
\end{equation}
yields the result equation (22).

\subsection{Long lime solution}
Laplace transform of the system takes the form 
\[ \left( \begin{array}{ccc}
s & 0 & 2 J \\
0 & s + \eta & -\Delta \\
-2 J & \Delta & s+\eta \end{array} \right)  \left( \begin{array}{c}
\tilde{r}(s)\\
\tilde{x}(s)\\
\tilde{y}(s) \end{array} \right)  =  \left( \begin{array}{c}
1\\
0\\
- J/s \end{array} \right)\] 
After inversion, in the long time limit, one has
\begin{equation}
r(t) \simeq 1/2 + rational~function(J, \Delta, \eta) e^{-t4 \frac{\eta J^2}{\eta^2 + J^2 + \Delta^2}}\nonumber
\end{equation}
This takes the value $1/2$ when $t>> t_{relax}= \frac{\eta^2 + J^2 + \Delta^2}{4 \eta J^2}$.

\subsection{Brief introduction to 2D photon echo spectroscopy}

A brief overview of 2D photon echo spectroscopy is given with an emphasis on the underlying  physics of
multidimensional echo spectroscopy (see for details\cite{2dphoton}). 2D photon echo spectroscopy (2DPES) is a
kind of generalization of the Pump-Probe Spectroscopy (PPS). In optical PPS, an ultra-short pump pulse with wide
bandwidth creates excitation of various electronic transitions and the subsequent probe pulse selectively
measures the transient absorption of the electronic states. This transient probe absorption is a function of
 delay time between the pump and the probe pulses. Thus one can get dynamical
information (temporal changes of absorption) of the relaxation processes. But the pump-probe spectroscopy is
insensitive to the coherences created by the optical
excitation whereas  2DPES is coherence sensitive, it can temporally resolve the dynamics of the coherence
(off-diagonal elements of the density matrix). In 2DPES three ultra short pulses are send through the sample.
The first pulse creates the coherence state between the ground and excited states$\rho_{ge}^{(1)}$. This evolves
for some  time period $\tau$ (order of femto seconds), then the next pulse interacts with this already excited
system. This yields either the ground state $\rho_{gg}^{(2)}$ or inter-exciton coherence state
$\rho_{ee'}^{(2)}$. This doubly excited state then evolves for another interval of time called population time
$T$ until a third pulse interacts with the system. This third interaction finally yields 3rd order density
matrix elements such as $\rho_{e'g}^{(3)}$ which emits an echo signal (in phase matched 
direction) by decaying after a time interval $t$. This can be used to measure real
time dynamics of resonance coupling in FMO systems and this can shed light on the conformal changes in
molecular structure such as hydrogen bound breaking (see below).

The usual linear spectroscopic methods like linear absorption or pump-probe spectroscopy can only provide highly
averaged information about the system under study, 
for example, in linear absorption spectra the broadening is both due to homogeneous broadening (HB) and
inhomogeneous broadening (IHB). But 2D photon echo spectroscopy can resolve these two contributions.

The way in which it resolves can be explained as follows. Consider that an ultra-short pulse perturbs the system
at time $t=0$. Consider that our system is composed of several
chromophors with different electronic transition frequency (static in-homogeneity) and it is interacting with
thermal bath (some  protein). The first pulse creates the 
coherence  state between the ground and excited state $\rho_{ge}^{(1)}$ of a choromophore  due to the dipolar
matrix elements coupling the ground and excited state (considering  that the 
light-matter  interaction is treated with first order perturbation theory--weak field regime). We have an
ensemble of coherences $\rho_{ge}$ with a specific phase relation 
at $t=0$.  Due to static in-homogeneity the phase relation between these density matrix 
elements will  be lost with time (phase randomness) but it will re-appear after sufficiently long time ! (if we
consider for the moment that there is no bath and no random 
perturbation of 
the phases).
Let this coherences $(\rho_{ge}^{(1)}(t))$ evolves for 
some  time period $\tau$ (order of femto seconds), then the next ultra short pulse interacts with this already
excited and evolving system. This yields either the ground state 
populations $\rho_{gg}^{(2)}$ (no coherences) or inter-exciton coherence states $\rho_{ee'}^{(2)}$.
These doubly excited states then evolves for another interval of time called population time $T$ until a third
pulse interacts with the system.  This third interaction has an opposite effect and creates the coherences
$\rho_{eg}^{(3)}$ (which are complex conjugates of the first coherences).
The time evolution of this exactly cancel the ``phase randomness`` developed in the initial time interval $\tau$
(because evolution operator for $\rho_{eg}$ is the complex
conjugate of the evolution operator for $\rho_{ge}$. If there is no bath (system is isolated) then after an
interval
of time $\tau$ the phases again "cohere" (they assume the same distribution as they had at time $t=0$) and 
finally this "re-locking" of phases yield an echo signal 
(in phase matched   direction).

Now consider that our system is interacting with the bath (our system is open). This cause a ''stochastic phase
randomness" between the phases of $\rho_{ge}$ of various 
chromophors in the ensemble.
If the population time $T$ is sufficiently long the phase relationship between density matrix elements  of the
chromophores will be {\it permanently} lost and and no echo 
will
be seen. Thus, we can say that the maximum population time $T_{max}$ directly depends upon (a) the strength of
system-bath coupling, (b) measure of the "fastness" of the bath 
fluctuations. In Kubo's stochastic theory these are respectively $\Delta$ and $\gamma$. Thus $T_{max}$ tell us
about the character of homogeneous broadening mechanism. The 
$\tau-\tau$ correlation or in frequency domain $\om_\tau -\om_\tau$ correlation (along the diagonal direction in
2D spectra) tell us about the in-homogeneous broadening, as 
the experiments are done by systematically varying $\tau$ and at a given $\tau$ (the time gap between the first
two pulses) the amplitude in the 2 D spectrum  after 
that given time $\tau$ from the second pulse  (i.e., after population time) show a correlation in the form of
the elongation of the  peak in the diagonal direction, 
a direct signature of in-homogeneous broadening. Thus one get the information about both HB and IHB.

 %-----------------------------------fig. --------------------------
\begin{figure}[!h]
\centering
\begin{tabular}{c}
\includegraphics[height = 8cm, width = 8cm]{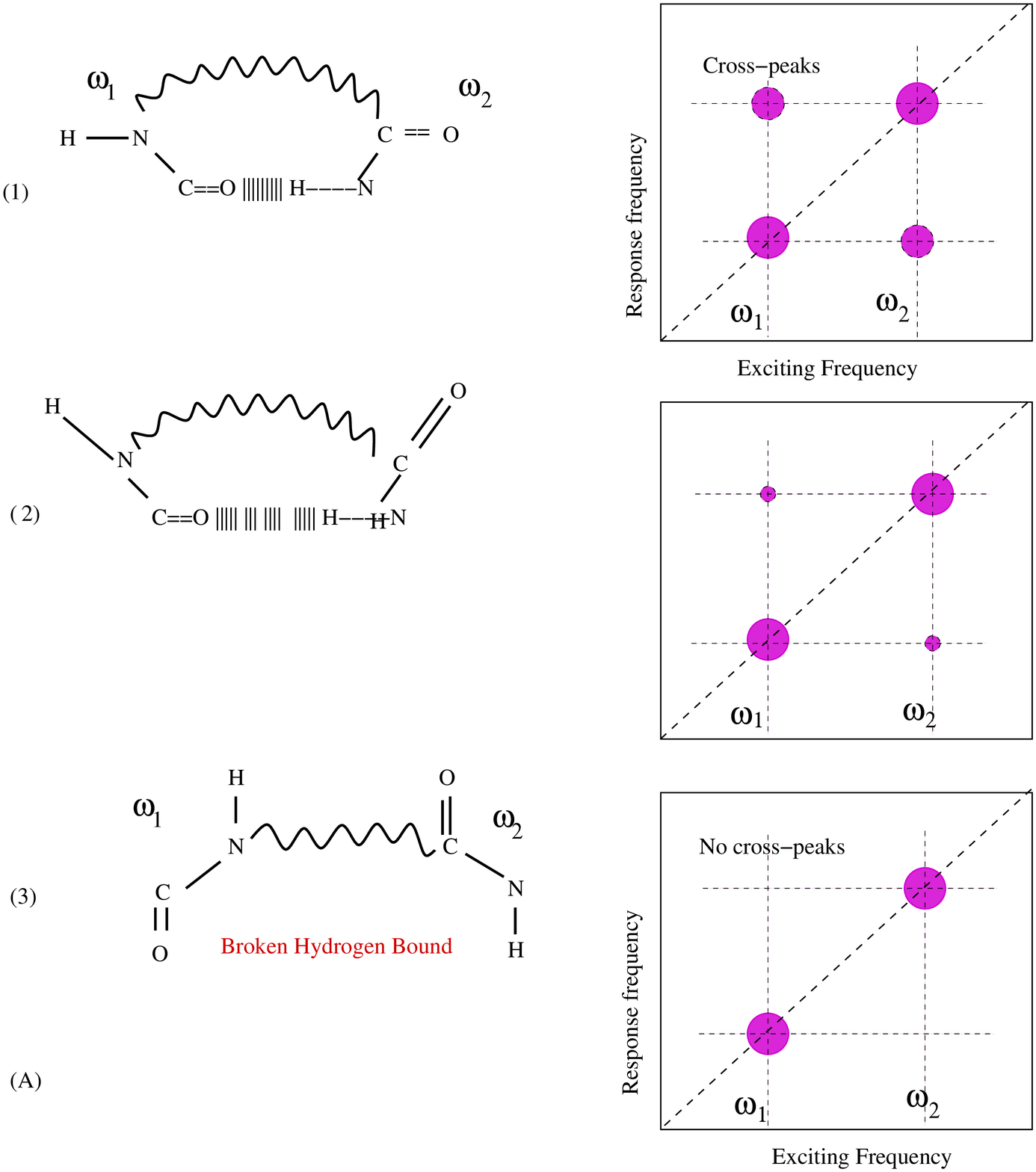}
\includegraphics[height = 8cm, width = 8cm]{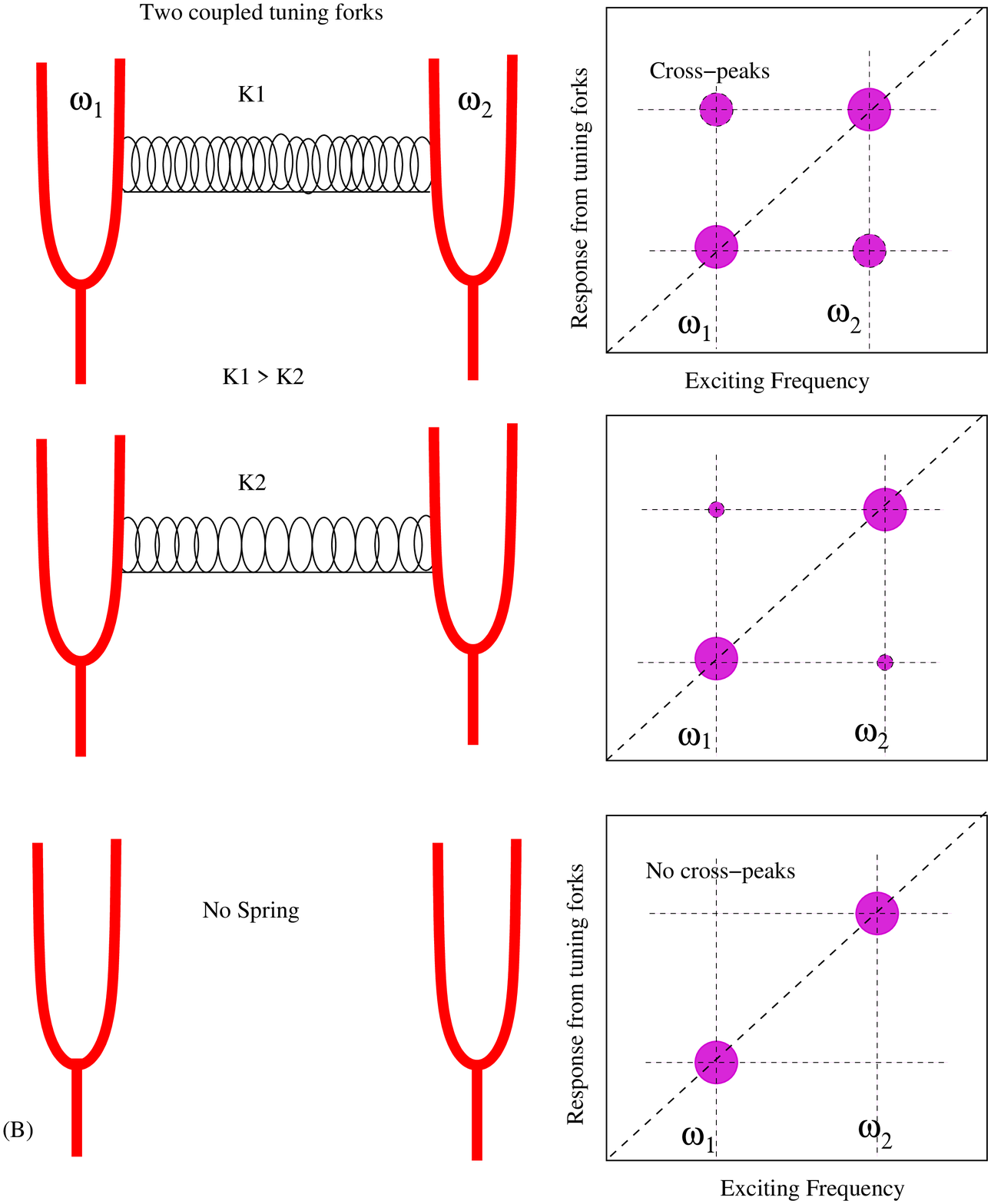}
\end{tabular}
\caption{(A) 2D spectroscopy can track (in real time) the changes in the molecular structure. In the above
example hydrogen bound breaking dynamics 
between two chemical groups is shown. Two chemical groups have characteristic frequencies $\om_1$ and $\om_2$.
Two peaks appear at these frequencies in the 
2D photon echo spectrum. Two cross-peaks also appear as shown which is due to the coupling of these two chemical
groups. The coupling is due to the hydrogen bound.
When the bound is broken this cross-peaks also disappear. (B) A mechanical analogy: The occurrence of cross
peaks in the exciting-response frequency spectrum 
is due to the spring coupling between two tuning  forks of frequency $\om_1$ and $\om_2$. The amplitude of
cross-peaks in the 2D spectrum measures the strength of 
coupling. When the spring constant decreases the cross peaks diminishes and finally goes to zero with no
coupling.}\label{tuning}
\end{figure}
%----------------------------------------------------------------------

2D photon echo spectroscopy can also tell us about  the real time  resonance coupling dynamics and bond breaking
dynamics (Fig.~\ref{tuning}). Consider that we have two tuning forks with 
 characteristic  frequencies $\om_1$ and $\om_2$. Let us excite this system with a spectrum of frequencies and 
detect the amplitude of vibration with some frequency
 analyzer (some electronic instrument) and plot various frequencies along the x-axis (exciting frequencies) and
y-axis (detecting frequencies). We will see two peaks 
occurring  at $\om_1$ and $\om_2$  in the "2D spectrum" (along the diagonal Fig.~\ref{tuning}).

Now consider that our tuning forks are coupled by some spring (say we have weak coupling). Then again repeat the
experiment and plot the 2D spectrum. This time we will see,
 along with the diagonal peaks, two "cross-peaks" along the anti-diagonal direction. These cross peaks are the
consequence of coupling. If we further reduce the spring 
coupling the magnitude of these cross peaks diminish and finally disappear with our  removal of  coupling
springs.

This mechanical analogy can be directly applied to the changes in the molecular structure. In  Fig. 1 hydrogen
bound breaking dynamics 
between two chemical groups is shown. Two chemical groups have characteristic frequencies $\om_1$ and $\om_2$.
Two peaks appear at these frequencies in the 
2D photon echo spectrum. Two cross-peaks also appear as shown which is due to the coupling of these two chemical
groups. The coupling is due to the hydrogen bound.
When the bound is broken this cross-peaks also disappear. This can be tracked in real time by varying the
magnitude of the population time. Similarly  in the electronic
spectrum the cross peaks tell us about the electronic coupling between the chromophores (the system-system
coupling, usually denoted as $J$ or $V$) also called 
resonance coupling. The oscillation in the magnitude of the cross peaks show the oscillations of the pigments
about their equilibrium positions (in physical space) 
as the variation about  the equilibrium position modulate the resonance coupling strength.

A brief mathematical formulation can  be described as follows. In semi-classical approximation for field-matter
interaction 
the interaction Hamiltonian is written as

\begin{equation}
 H_{int} = - {\bf\mu}. {\bf E}({\bf r},t)
\end{equation}
 
Here  dipole approximation (weak variation of the electric field amplitude over the size of the pigment) is
used. If the magnitude of the above Hamiltonian is much 
weaker than the magnitude of the pigment Hamiltonian $H_{pig} = H_e + H_{ph} + H_{e-ph}$, then the $H_{int}$ can
be treated as a perturbation. 

The dynamics of the total system (the pigment) can be described by Liouville-von Neumann equation for the
density matrix $\rho$.
\begin{equation}
\frac{\pr\rho}{\pr t} = -\frac{i}{\hbar} [H_{pig} + H_{int},\rho]
\end{equation}
Now consider that the sample is interrogated with three consecutive (in time) ultra short laser pulses. Treating
$H_{int}$ as a perturbation
the third order density matrix is given as
\begin{eqnarray}
&&\rho^{(3)}({\bf r},t) = (\frac{i}{\hbar})^3 \int_0^\infty dt_3  \int_0^\infty dt_2   \int_0^\infty dt_1
\Theta(t_3) e^{-\frac{i}{\hbar} \ml t_3}\times \\
&& L_\mu \Theta(t_2) e^{-\frac{i}{\hbar} \ml t_2} L_\mu \Theta(t_1) e^{-\frac{i}{\hbar} \ml t_1} L_\mu
\rho(-\infty) {\bf E}({\bf r},t-t_3) {\bf E}({\bf r},t-t_3-t_2)
{\bf E}({\bf r},t-t_3-t_2-t_1)\nonumber.
\end{eqnarray}
Here $\ml * = [H_{pig}, *] $ and $L_\mu * = [{\bf \mu},*]$.  Experimentally we do not measure $\rho(t)$ but the
induced polarization $P^{(3)}({\bf r},t) = tr[\mu \rho^{(3)}(t)]$.

\begin{equation}
{\bf P}^{(3)}({\bf r},t) = (\frac{i}{\hbar})^3 \int_0^\infty dt_3  \int_0^\infty dt_2   \int_0^\infty dt_1 
R^{(3)}(t_3,t_2,t_1) {\bf E}({\bf r},t-t_3) 
{\bf E}({\bf r},t-t_3-t_2)
{\bf E}({\bf r},t-t_3-t_2-t_1).
\end{equation}

The 3rd order response function is given as
\begin{equation}
 R^{(3)}(t_3,t_2,t_1)= (\frac{i}{\hbar})^3 \Theta(t_3)\Theta(t_2)\Theta(t_1) \left\la
[[[\mu(t_3+t_2+t_1),\mu(t_2+t_1)],\mu(t_1)],\mu(0) \rho(0)]\right\ra.
\end{equation}

The time dependence of $\mu$ can be shifted to the time dependence of the density matrix using time evolution
operators (resulting $\mu$ as time independent). Ultimately one have to take the trace over the bath modes thus
one will end up having an expression involving time depended reduced density matrix 
elements of the system.

The electric field of the incoming pulses can be given as

\begin{equation}
 {\bf E}({\bf r},t) = \sum_{i=1}^3 \left({\bf n_i} E_i(t) e^{(i.{\bf k_i.r} - i \om_i t)} + c.c.\right)
\end{equation}

Where ${\bf n}$ is the unit vector pointing in the direction of the field and $E(t)$ is the temporal envelope 
of the pulse. The non-linear polarization can also be 
expanded in the Fourier components

\begin{equation}
 {\bf P}^{(3)}({\bf r},t) = \sum_l  {\bf P}^{(3)}_l(t) e^{i{\bf k_l.r} - i \om_l t } 
\end{equation}
Where

\begin{equation}
 {\bf k}_l = \pm {\bf k}_1 \pm   {\bf k}_2 \pm  {\bf k}_3,~~~~ \om_l = \pm \om_1 \pm \om_2 \pm \om_3
\end{equation}

This non-linear polarization will emit a signal electric field in the phase matching direction. It can be
shown\ref{}( if the electric field amplitude varies slowly over the pigment size and if the refractive  index of
the material is frequency independent) that 
\begin{equation}
 {\bf E}_{signal}^{(3)} \propto  {\bf P}^{(3)}
\end{equation}
Now under the impulsive limit (excitation femtosecond laser pulses as delta functions)
$E_1(t)=E_1\delta(t+T+\tau),~~ E_2(t)= E_2\delta(t+T),~~E_3(t)=E_3\delta(t)$ the
 signal electric field becomes proportional to the response  function. The Double Fourier transform of the
signal field gives the complex 2D spectrum
\begin{equation}
 \tilde{E}_{signal}^{(3)} (\om_t,T,\om_\tau) \propto \int_{-\infty}^{+\infty} dt  \int_{-\infty}^{+\infty} d\tau
R^{(3)}(t,T,\tau) e^{i \om_t t} e^{i\om_\tau \tau}
\end{equation}
 
This is the 2D photon echo signal. The whole problem boils down to the calculation of 3rd order response
function which further involves the time evolution of the reduced density matrix.

%************************************************************


\begin{thebibliography}{100}
\bibitem{note100} First, why bath should re-organize? The answer is: when an electronic transition happens it
disturbes the charge distribution in the molecule (pigment). Nuclear co-ordinates re-organize in real space to
reach a new equilibrium configuration.
\end{thebibliography}
\end{document}